\documentclass[twocolumn,showpacs,preprintnumbers,amsmath,amssymb]{revtex4}

\usepackage{graphicx}
\usepackage{dcolumn}
\usepackage{bm}
\usepackage[colorlinks]{hyperref}

\begin{document}


\title{Traffic properties for stochastic routings on scale-free networks}

\author{Yukio Hayashi}
\email{yhayashi@jaist.ac.jp}
\author{Yasumasa Ono}
\affiliation{
Japan Advanced Institute of Science and Technology,\\
Ishikawa, 923-1292, Japan
}

\date{\today}

\begin{abstract}
For realistic scale-free networks, 
we investigate the 
traffic properties of stochastic routing inspired by
a zero-range process known in statistical physics.
By parameters $\alpha$ and $\delta$,
this model controls degree-dependent hopping of packets 
and forwarding of packets with higher performance at more busy nodes.
Through a theoretical analysis and numerical simulations,
we derive the condition for the 
concentration of packets at a few hubs.
In particular, we show that the optimal $\alpha$ and $\delta$
are involved in the trade-off between 
a detour path for $\alpha < 0$ 
and long wait at hubs for $\alpha > 0$; 
In the low-performance regime at a small $\delta$,
the wandering path for $\alpha < 0$ better reduces the 
mean travel time of a packet with high reachability.
Although, in the high-performance regime at a large $\delta$,
the difference between $\alpha > 0$ and $\alpha < 0$ is small,
neither the
wandering long path with short wait trapped at nodes
($\alpha = -1$),
nor the short hopping path
with long wait trapped at hubs ($\alpha = 1$) is advisable.
A uniformly random walk ($\alpha = 0$)
yields slightly better performance.
We also discuss the congestion phenomena in a more complicated situation
with packet generation at each time step.
\end{abstract}

\maketitle


\section{Introduction}
 In daily socio-economic network systems, 
many commodities, passengers, or
information fragments (abstractly referred to as 
{\em packets}
in this paper) are delivered from one place to another place.
In general, it is expected to send and receive packets as quickly as
possible without encountering traffic congestion 
which would force packets to wait at nodes.
Even if packets are concentrated (or {\em condensed} in term of
statistical physics)
on just a few nodes in some parts of a network, 
this situation may cause congestion over the network. 
Thus, one of the important
issues is a routing strategy: how to select a forwarding node
in the neighbors of the resident node of a packet. 
Since the efficiency of transportation or communication depends 
on not
only routing strategy, but also on network topology, we should
consider a realistic problem setting for the routing 
and the topology.
In addition, 
considering the interaction of accumulated packets 
in a buffer (called 
{\em queue}) at each node is necessary, 
since it crucially affects the traffic flow on a network.
This paper discusses 
traffic properties that depend on routing strategies 
in the interaction of packets on a realistic network topology.

During this decade, a new research field, 
{\em complex network science},
has been created \cite{Albert02}, 
since a common topological structure called {\em scale-free}
(SF) was found to exist 
in many real networks such as the Internet, the
World-Wide-Web, power grids, airline networks, 
social collaboration networks, human sexual contacts, 
and biological metabolic pathways, etc. 
The SF structure is characterized by the property that the
distribution of degree (the number of connections to a node) 
follows a power-law. 
In other words, the network consists of many nodes with low degrees
and a few hubs with high degrees. 
Moreover, a SF network naturally emerges in 
social acquaintance relationships or peer-to-peer communications, 
and it has short paths compared with large network size 
(the total number of nodes). 

In a realistic SF network, 
a routing path is shortened by passing through hubs 
for the forwarding of a packet.
However, many packets may be concentrated at the hubs \cite{Goh01}, 
in whose queues the packets are cumulatively stored 
if they arrive in a quantity larger than the processing limit
for forwarding. 
In such a situation, 
there exists a trade-off between delivering packets on a short path 
and avoiding traffic congestion. To improve 
communication efficiency even in a dynamic environment for a wireless
or ad hoc network, various routing schemes are being developed. 
Because, in an ad hoc network, 
many nodes (such as base stations or communication sites) 
and connections between them 
are likely to change over time, 
then global information, e.g., a routing table in the Internet,  
cannot be applied.
In early work, 
some decentralized routing developed to reduce energy
consumption in sensor or mobile networks, 
however they lead to the failure of
guaranteed delivery \cite{Urrutia02};
in the flooding algorithm, multiple redundant
copies of a message are sent and cause congestion, while
greedy and compass routing may occasionally fall into infinite loops
or into a dead end.
Thus, we focus on 
stochastic routing methods using only local information with
respect to the resident node of a packet and 
to the connected neighbors on a path, 
due to their simplicity and power.
If the terminal node of a packet is included in the neighbors of the 
resident node, 
then the packet is 
deterministically sent to the terminal in order to guarantee 
reachability unless the connectivity is broken. 
We call this neighbor search (or n-search)
only at the last step to the terminal.
Note that the development of stochastic routing methods is
still in progress depending on device and information
processing technologies.
Although we suppose a mixture 
of wireless and wired communications in
future networks, first of all, 
we aim to understand the relations for traffic properties
between fixed network topology and routing methods.
Here, it is usually assumed for simplicity 
that each node has the 
same performance in the forwarding of packets.

  The enhancement of performance in 
forwarding is also reasonable \cite{Liu06,Zhao05}.
For example, in the
Internet or airline networks, an important facility with many
connections has high performance; more packets or flights are
processed as the incoming flux 
of communication requests or of passengers
increases. Such higher performance at more busy nodes is effectively
applied in a {\em zero-range process} (ZRP) \cite{Noh05a,Noh05b},
which tends to distribute packets throughout 
the whole network, instead of avoiding paths through hubs.

In this paper, from the random walk version, we extend the ZRP on a
SF network to the degree-dependent hopping rule,
which parametrically controls routing strategies in the trade-off
between the selection of a short path passing through hubs, 
and the avoidance of hubs 
at which many packets are condensed and 
wait in queues for a long time. 
The above models in statistical physics are useful 
for the analysis of complex phenomena involved in transition 
form a free-flow 
phase to a congestion phase, or the opposite transition.
Most research on traffic congestion rely on numerical
simulations.
Although a few other theoretical analyses based on the mean-field
approximation \cite{Martino09} have been done, we can not compare 
them to the ZRP, simply because of different problem settings. 
Thus, we consider a combination of settings as modified traffic
models, and numerically investigate them.

The organization of this paper is as follows.
In Sec. \ref{sec2}, we briefly review the related models in recent
studies.
In Sec. \ref{sec3}, we introduce a stochastic 
packet transfer model defined by the ZRP, 
in which all packets persist without any generation and removals.
Then, in terms of fundamental properties, 
we approximately analyze 
the stationary probability of incoming packets at each node 
on SF networks, 
and derive the phase transition for condensation of packets at hubs 
in the degree-dependent hopping rule.
In Sec. \ref{sec4}, 
we numerically confirm 
the phase transition, and as a new result, 
show the trade-off between a
detour wandering path and long wait at hubs. 
Moreover, 
we discuss 
the congestion phenomenon in the modified traffic models 
with packet generation.
In Sec. \ref{sec5}, we summarize these results
and describe some issues for further research.

\section{Related Work} \label{sec2}
Many traffic models have been proposed in various problem settings
for routing, node capacity, and packet generation.
They are summarized in Table \ref{table_models}. 
The basic processes for packet transfer consist of 
the selection of a forwarding (coming-in) node 
and the jumping-out of packets in the node capacity. 
In these models (including our model discussed later), 
a significant issue commonly arises from 
the trade-off between delivering packets on a
shorter path and avoiding the congestion caused by 
a concentration of packets on a few nodes such as hubs.

In the typical models,  
a forwarding node $k$ is chosen with probability either 
$K_{k}^{\alpha} / \sum_{j \in {\cal N}_{i}} K_{j}^{\alpha}$ 
\cite{Yamashita03,Wang06a,Yan06} 
or 
$(m_{(k)} + 1)^{- \beta} / \sum_{j \in {\cal N}_{i}} (m_{(j)} +
1)^{- \beta}$  \cite{Danila06}.
Here, $\alpha$ and $\beta$ are real parameters, 
$K_{k}$ and $m_{(k)}$ denote the degree
and the dynamically-occupied queue length by packets 
at node $k$ in the connected neighbors 
${\cal N}_{i}$ to the resident node $i$ of the packet. 
These methods are not based on a random walk 
(selecting a forwarding node 
uniformly at random among the neighbors), 
but on the extensions (including the uniformly random one 
at $\alpha, \beta = 0$)
called preferential and congestion-aware walks, respectively. 
Note that  $\alpha > 0$ leads to  a short path passing 
through hubs, 
and that $\alpha < 0$ and $\beta > 0$ lead to the avoidance of hubs
and congested nodes with large $m_{(k)}$.
In stochastic routing methods, instead of using the shortest path, 
the optimal values 
$\alpha = -1$ and $\beta = 1$ for maximizing 
the generation rate of packets in a free-flow regime 
have been obtained by numerical simulations \cite{Wang06a,Yan06}. 
A correlation between the congestion at node level and a betweenness 
centrality measure was suggested  \cite{Danila06}.

Other routing schemes \cite{Wang06b,Echenique05,Martino09} 
have also been considered, 
taking into account lengths of both the routing path and of the queue.
In a deterministic model \cite{Echenique05}, 
a forwarding node $k$ is chosen among neighbors ${\cal N}_{i}$
by minimizing the quantity $h d_{k} + (1 - h) m_{(k)}$ 
with a weight $0 \leq h \leq 1$, 
$d_{k}$ denoting the distance from $k$ to the terminal node.
Since we must solve the optimization problems, 
these models \cite{Yan06,Echenique05} are not suitable for wireless 
or ad hoc communication networks.
Thus, stochastic routing methods using only local information
are potentially promising.
In a stochastic model \cite{Martino09}, 
$k \in {\cal N}_{i}$
is chosen at random, and 
a packet at the top of its queue is sent  
with probability $1 - \eta(m_{(k)})$
or refused  
with probability $\eta(m_{(k)})$ as a nondecreasing function of  
the queue length $m_{(k)}$.  
This model is simplified by the assumption of a constant 
arrival rate of
packets, for analyzing the critical point of traffic congestion in a
mean-field equation \cite{Martino09}.

With a different processing power at each node \cite{Wang06a}, 
it has also been considered that 
the node capacity $c_{i}$ 
is proportional to its degree $K_{i}$,
therefore more packets jump out from a node as the
degree becomes larger.
On the other hand, 
in the ZRP \cite{Noh05a,Noh05b,Noh07}, 
the forwarding capacity at a node depends on the number of $m_{(i)}$
defined as a queue length occupied by packets at node $i$.
The ZRP is a solvable theoretical model for traffic dynamics. 
In particular, in the ZRP with a random walk at $\alpha = 0$, 
the phase transition between condensation of packets at hubs 
and uncondensation on SF networks 
has been derived \cite{Noh05a,Noh05b}. 
For $\alpha > 0$, 
a similar phase transition has been analyzed 
in the mean-field approximation \cite{Tang06}.

In the next two sections, 
based on a straightforward 
approach introduced in Refs. \cite{Noh05a,Noh05b}, 
we derive the phase transition
in the ZRP on SF networks with the degree-dependent hopping rule 
for both $\alpha > 0$ and $\alpha < 0$, inspired by
preferential \cite{Yamashita03,Wang06a,Yan06} 
and congestion-aware walks. 
Although the rule is not identical to 
the congestion-aware routing scheme \cite{Danila06}
based on occupied queue length $m_{(k)}$, 
$\alpha < 0$ corresponds to avoiding hubs with large degrees, 
where many packets tend to be concentrated.
Furthermore, we study the traffic properties in the case with 
neighbor search into a terminal node 
at the last step.

\begin{table}[htb]
\caption{Recent traffic models for complex networks. 
When a packet is forwarded 
from a current node $i$ to $k \in {\cal N}_{i}$, 
the neighboring node $k$ is chosen 
with probability $\Pi_{k}$ 
or by minimizing an objective function on a routing path.
The node capacity $c_{i}$ is defined by 
the number of simultaneously transferable packets 
from each node $i$.
Here, 
$K(x_{l})$ denotes the degree of node $x_{l}$, 
$\Theta(x)$ is the step function, and 
$m^{*}$ is a threshold, $\beta \geq 0$, and 
$0 \leq \bar{\eta} < 1$.}
\begin{center}
\begin{footnotesize}
\begin{tabular}{c|ccc} \hline\noalign{\smallskip}
Ref. & selection of & node & packet \\
     & forwarding node 
               & capacity & generation \\  \noalign{\smallskip} \hline
 \noalign{\smallskip}
 \cite{Wang06a} & $\Pi_{k} \propto K_{k}^{\alpha}$ 
                                  & $c_{i} = 10$ or & Yes \\
               & $-1 \leq \alpha \leq 1$ & $c_{i} = K_{i}$ & \\ \hline
 \cite{Yan06}  & $\min \; \sum_{l} K(x_{l})^{\beta}$
                                  & $c_{i} = 1$ & Yes \\ 
               & on a path $\{x_{0}, \ldots, x_{n} \}$ & & \\ \hline
 \cite{Danila06} & $\Pi_{k} \propto (m_{(k)}+1)^{-\beta}$ 
                                  & $c_{i} = 1$ & Yes \\ 
                 & & & \\ \hline
 \cite{Wang06b} & $\Pi_{k} \propto K_{k}(m_{(k)}+1)^{-\beta}$ 
                                  & $c_{i} = 5$ & Yes \\ 
                 & & & \\ \hline
 \cite{Echenique05} & $\min \; h d_{k} + (1-h) m_{(k)}$
                                  & $c_{i} = 1$ & Yes \\ 
                    & $0 \leq h \leq 1$ & & \\ \hline
 \cite{Martino09} & random walk & & \\
                  & with a refusal prob. & $c_{i} = 1$ & Yes \\ 
                  & $\bar{\eta} \Theta(m_{(k)} - m^{*})$ & & \\ \hline
 \cite{Noh05a,Noh05b,Noh07}
                 & $\Pi_{k} \propto K_{k}^{\alpha}$ 
                                  & jumping rate & No \\
 \cite{Tang06} & $\alpha > 0$     & $m_{(i)}^{\delta}$ & 
 \\ \noalign{\smallskip} \hline \noalign{\smallskip}
\end{tabular}
\end{footnotesize}
\end{center}
\label{table_models}
\end{table}

\section{Packet Transfer Model} \label{sec3}
\subsection{Routing rule}
Consider a system of $M$ interacting packets 
on a network of $N$ nodes with a density $\rho = M/N$,  
$M = \sum_{i=1}^{N} m_{(i)}$, 
where $m_{(i)} \geq 0$ denotes 
the occupation number of packets in the queue 
at each node $1 \leq i \leq N$. 
For simplicity, 
we assume that the queue length is not limited, 
and that the order of stored packets is ignored. 
If there is a limitation on the queue length, 
it may become necessary to discuss a 
 cascading problem \cite{Motter02,Motter04,Wu07} 
whose dynamics are very complicated. 

To investigate the predicted properties from 
the theoretical analysis in the ZRP \cite{Noh05a,Noh05b,Noh07}, 
we use the same problem setting, such 
that all packets persist on paths in the network
without generation or removals of packets. 
For more complex situations 
with packet generation, 
some models modified by adding different routing rules 
will be investigated in subsection 4.2.

In this routing rule related to the ZRP, 
the total number $M$ of packets 
is constant at any time, and 
each node performs a stochastic local search as follows: 
at each time step, 
a packet jumps out of a node $i$ stochastically 
at a given rate $q_{i}(m_{(i)})$
as a function of $m_{(i)}$, 
and then comes into one of the neighboring nodes $k \in {\cal N}_{i}$ 
chosen with probability
$K_{k}^{\alpha} / \sum_{k' \in {\cal N}_{i}} K_{k'}^{\alpha}$. 
When a packet is transferred from $i$ to $k$, 
the queue length $m_{(k)}$ is increased by one unit, 
and $m_{(i)}$ is simultaneously decreased.
The above processes are conceptual, 
the relaxation dynamics for the simulation of packet transfer 
is necessary, and described in  Sec. \ref{sec4}. 
The effect of the deterministic neighbor search 
into a terminal node at the last step will also be discussed later.

\subsection{Stationary probability in 
$\alpha$-random walks}
Before investigating the influence of 
dynamic queue lengths on traffic properties, 
we consider 
the stationary probability of incoming packets at each node.
As mentioned in the previous subsection, 
we extend a random walk routing \cite{Noh07,Noh04},  
in which a walker (packet)
chooses a node uniformly at random among the neighbors of
the current node on a path,  
to a degree-dependent routing. 
We call it $\alpha$-random walk.

\begin{figure}[htb]
 \begin{center}
  \resizebox{8cm}{!}{
    \includegraphics{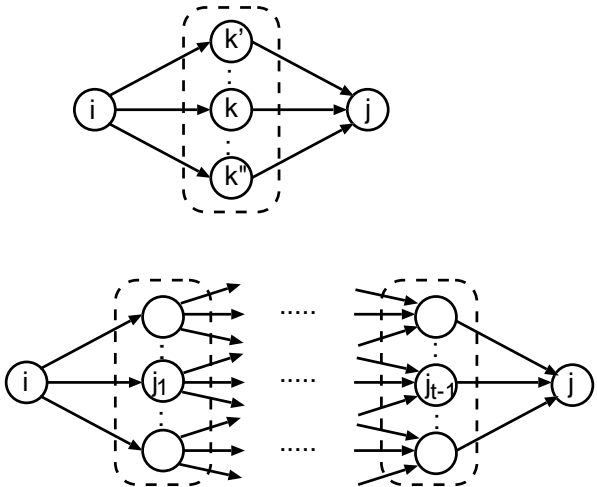}}
  \caption{Combination of consecutive nodes 
  on the route from node $i$ to $j$. 
  The dashed rectangles correspond to the sum 
  of Eqs. (\ref{eq_alpha_rand})-(\ref{eq_Pji}).} 
  \label{fig_route_from-to}
 \end{center}
\end{figure}

As in Ref. \cite{Noh04}, 
for the probability $P_{ij}$ of finding the walker
at node $j \in {\cal N}_{k}$ from node $i$ 
through the intermediate nodes $k \in {\cal N}_{i}$ at time $t$,
the master equation is 
\begin{equation}
  P_{ij}(t+1) = \sum_{k} \frac{
  K_{j}^{\alpha}}{\sum_{j' \in {\cal N}_{k}}
  K_{j'}^{\alpha}} P_{ik}(t).
  \label{eq_alpha_rand}
\end{equation}
By iterating Eq. (\ref{eq_alpha_rand}), 
an explicit expression for the transition probability $P_{ij}$ 
to go from node $i$ through $j_{1}, \ldots, j_{t-1}$ 
to $j$ in $t$ steps follows as 
\begin{equation}
  P_{ij}(t) = \sum_{j_{1},  \ldots, j_{t-1}}
  \frac{K_{j_{1}}^{\alpha}}{\sum_{j'_{1} 
  \in {\cal N}_{i}} K_{j'_{1}}^{\alpha}}
  \times \ldots \times 
  \frac{K_{j}^{\alpha}}{\sum_{j' \in {\cal N}_{j_{t-1}}}
  K_{j'}^{\alpha}}, 
  \label{eq_Pij}
\end{equation}
where the sum $\sum_{j_{1},  \ldots, j_{t-1}}$
is taken over the connected paths between nodes $i$ and $j$,
as shown in Fig. \ref{fig_route_from-to}.
In the opposite directions of the same paths, 
the transition probability $P_{ji}$ follows as
\begin{equation}
  P_{ji}(t) = \sum_{j_{t-1}, \ldots, j_{1}}
  \frac{K_{j_{t-1}}^{\alpha}}{\sum_{j'_{t-1} \in {\cal N}_{j}}
  K_{j'_{t-1}}^{\alpha}} \times \ldots \times 
  \frac{K_{i}^{\alpha}}{\sum_{i'  \in {\cal N}_{j_{1}}}
  K_{i'}^{\alpha}}.
  \label{eq_Pji}
\end{equation}
We assume the network to be uncorrelated: 
there is no correlation  
between two degrees of any connected nodes, 
so that 
$\sum_{j'_{1}} K_{j'_{1}}^{\alpha} 
= K_{i} \cdot \langle K^{\alpha} \rangle$ 
and 
$\sum_{j'_{t-1}} K_{j'_{t-1}}^{\alpha}
= K_{j} \cdot \langle K^{\alpha} \rangle$ 
hold by using the mean value $\langle K^{\alpha} \rangle$
of the $\alpha$-power of degree. 
By comparing the expression of $P_{ij}$ in Eq. (\ref{eq_Pij})
with that of $P_{ji}$ in Eq. (\ref{eq_Pji}), 
we obtain the equivalent relation
\[
   \frac{K_{i}}{K_{j}^{\alpha}} P_{ij} 
   =       \frac{K_{j}}{K_{i}^{\alpha}} P_{ji}.
\]
Thus, for any source $i$ and step $t$, 
$P_{ij}$ is proportional to 
$K_{j}^{1 + \alpha}$, 
while $P_{ji}$ is proportional to  $K_{i}^{1 + \alpha}$. 
Consequently, 
the stationary solution $P_{j}^{\infty}$ 
of the incoming probability $P_{ij}$
is proportional to 
$K_{j}^{1 + \alpha}$.
This form, related only to the degree of the forwarding node $j$,
is suitable for the theoretical analysis of the ZRP presented 
in the next subsection and in the Appendix. 
The approximative solution is also obtained 
from a different approach to the master equation \cite{Wang06a} 
under the same assumption of uncorrelated networks. 

\begin{figure}[htb]
 \begin{center}
  \resizebox{9cm}{!}{
    \includegraphics{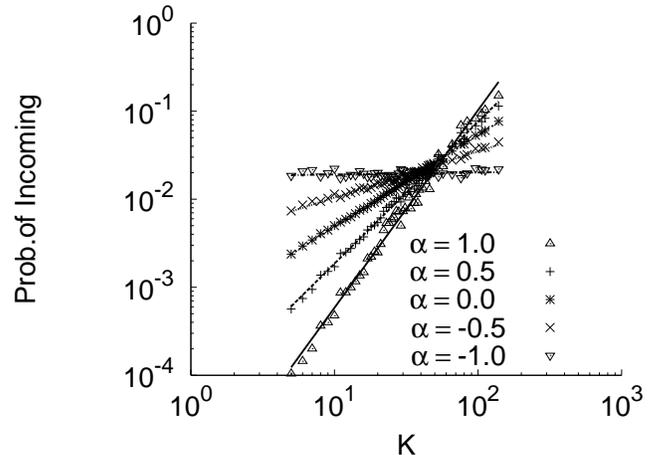}}
  \caption{Probability of incoming at a node
  with degree $K$ 
  in the $\alpha$-random walk of $M = 1000$ independent packets 
  (equivalent to the case of $\delta = 1$) through 1000 rounds. 
  The lines guide the estimated slopes 
  in the left column (without n-search) in Table \ref{table_est_beta}, 
  and the plotted marks
  show the probability obtained in a BA network with $N = 1000$ 
  and $\langle K \rangle = 10$.} 
  \label{fig_correl}
 \end{center}
\end{figure}

Figure \ref{fig_correl} shows the stationary solution 
$P_{j}^{\infty} \propto K_{j}^{\beta}$ and the exponent 
$\beta \approx 1 + \alpha \geq 0$ for 
the SF networks (generated by the 
BA: Barab\'{a}si-Albert model \cite{Albert02}).
Table \ref{table_est_beta} also shows that 
the estimated exponents $\beta$ are consistent in the cases both
with and 
without neighbor search into a terminal node at the last step.
Here, the terminal is chosen from all nodes uniformly at random.
After arriving at the terminal, the packet is restarted 
(resent) from the node to a new terminal,
in order to maintain the persistency of packets in the ZRP.

\begin{table}[htb]
\caption{Numerically estimated exponent $\beta \approx 1 + \alpha$
 for the lines from the plotted marks in Fig. \ref{fig_correl} 
 by using the mean-square-error method 
 for the independent walks of $M = 1000$ packets.
 These values give the averaged slopes for 
 $P_{j}^{\infty} \propto K_{j}^{\beta}$ in 
 100 BA networks with $N = 1000$ and $\langle K \rangle = 10$.}
\begin{center}
\begin{footnotesize}
\begin{tabular}{c|cc} \hline\noalign{\smallskip}
         & without & with n-search \\ 
$\alpha$ & $\beta$ & $\beta$       \\  \noalign{\smallskip} \hline \noalign{\smallskip}
 1.0     & 2.245   & 2.143 \\
 0.5     & 1.611   & 1.576 \\
 0.0     & 1.039   & 1.027 \\
-0.5     & 0.524   & 0.519 \\
-1.0     & 0.022   & 0.019 \\ \noalign{\smallskip} \hline \noalign{\smallskip}
\end{tabular}
\end{footnotesize}
\end{center}
\label{table_est_beta}
\end{table}

\subsection{Phase transition in the ZRP}
We discuss the phase transition between 
condensation and uncondensation
in the ZRP with degree-dependent hopping rule of packets. 
For the configuration 
\[
 m_{(1)}, \ldots, m_{(i)}, \ldots, m_{(N)}
\]
of occupation at each node, 
the stationary solution of a single packet 
is given by the factorized form 
\begin{equation}
   P(m_{(1)}, \ldots, m_{(i)}, \ldots, m_{(N)}) 
   = \frac{1}{Z} \Pi_{i=1}^{N} f_{i}(m_{(i)}), 
   \label{eq_P_factorize}
\end{equation}
where $Z$ is a normalization factor and 
\begin{equation}
  f_{i}(m_{(i)}) \stackrel{\rm def}{=} 
  \Pi_{\omega = 1}^{m_{(i)}} 
  \left( \frac{P_{i}^{\infty}}{q_{i}(\omega)} \right), 
  \label{eq_def_fi}
\end{equation}
for an integer $m_{(i)} > 0$ and $f_{i}(0)=1$.
We consider a function 
$q_{i}(\omega) = \omega^{\delta}$ with a parameter 
$0 \leq \delta \leq 1$
for forwarding performance. 
This means that a node works harder for transfer, 
as it has more packets in the queue
with larger $\omega$ and $\delta$.

\begin{figure}[htb]
 \begin{center}
  \resizebox{8.5cm}{!}{
    \includegraphics{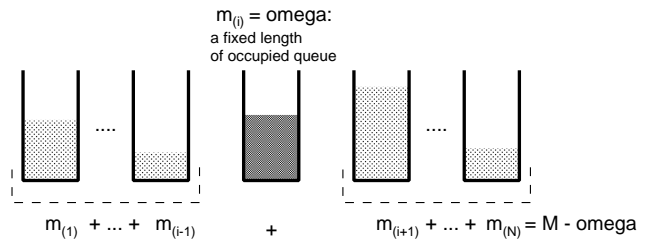}}
  \caption{Illustration of the queue length occupied by packets at each
  node. For the sum $\sum_{*}$ in Eq. (\ref{eq_Pi_omega}), 
  there are many combinations of 
  $\{ m_{(1)}, \ldots, m_{(i-1)}, m_{(i+1)}, \ldots, m_{(N)} \}$
  satisfying $\sum_{j \neq i} m_{(j)} = M - \omega$.} 
  \label{fig_combination_mK}
 \end{center}
\end{figure}

Using the probability distribution in Eq. (\ref{eq_P_factorize}) 
and the stationary probability $P_{i}^{\infty} \propto K_{i}^{\beta}$, 
we can calculate the mean value $\langle m_{(i)} \rangle$
at each node. 
Here, 
$\langle m_{(i)} \rangle \stackrel{\rm def}{=} 
\sum_{\omega=0}^{\infty} \omega P_{i}(m_{(i)} = \omega)$
is defined by using the probability distribution 
$P_{i}(m_{(i))} = \omega)$ of the number of packets occupying node $i$.
\begin{equation}
  P_{i}(m_{(i)} = \omega) = \frac{1}{Z_{i}} \sum_{*} f_{i}(\omega) 
  \Pi_{j \neq i} f_{j}(m_{(j)}), \label{eq_Pi_omega}
\end{equation}
where the sum $\sum_{*}$ is taken in combination 
$\{ m_{(1)}, \ldots, m_{(i-1)}, m_{(i+1)}, \ldots, m_{(N)} \}$
on the constraint $\sum_{j \neq i} m_{(j)} = M - \omega$ 
as shown in Fig. \ref{fig_combination_mK}, 
and $Z_{i}$ is a normalization factor. 
It is difficult to directly solve the normalization factor 
$Z_{i}$ or $Z$.
Thus, introducing a fugacity variable $z$ \cite{Evans00}, 
the mean value is given by a generating function 
\begin{equation}
  \langle m_{(i)} \rangle = 
  \frac{\sum_{\omega} \omega z^{\omega} 
           f_{i}(\omega)}{\sum_{\omega} z^{\omega} f_{i}(\omega)}
  =  z \frac{\partial \ln F_{i}(z)}{\partial z},
  \label{eq_mk_calc}
\end{equation}
where the second term in the right-hand side of Eq. (\ref{eq_mk_calc}) 
is due to the definition
\[
  F_{i}(z) \stackrel{\rm def}{=} 
  \sum_{\omega = 0}^{\infty} z^{\omega} f_{i}(\omega).
\]
From the definition (\ref{eq_def_fi}), 
$q_{i}(\omega) = \omega^{\delta}$, 
and $P_{i}^{\infty} \propto K_{i}^{\beta}$, 
we have 
\begin{equation}
  F_{i}(z) = \sum_{\omega = 0}^{\infty} 
   \frac{(z K_{i}^{\beta})^{\omega}}{(\omega \; !)^{\delta}},
   \label{eq_Fi_series}
\end{equation}
because 
$f_{i}(\omega) = \Pi_{m=1}^{\omega}
  \left( \frac{K_{i}^{\beta}}{m^{\delta}} \right)
  = \frac{(K_{i}^{\beta})^{\omega}}{(\omega !)^{\delta}}$
in Eq. (\ref{eq_def_fi}).
The fugacity $z$ should be determined from 
the self-consistency equation 
$\rho = \sum_{i=1}^{N} \langle m_{(i)} \rangle / N$
as a function of $z$. 
Note that 
$M = \sum_{i=1}^{N} \langle m_{(i)} \rangle$ is 
the total number of packets in a network of $N$ nodes,
and that the density $\rho$ is constant at 
$N, M \rightarrow \infty$.

In this paper,  
we consider a SF network
whose degree distribution 
follows a power-law $P(K) \sim K^{-\gamma}$. 
According to the performances of jumping-out, 
we classify the following cases 
(A) $\delta > \delta_{c} \stackrel{\rm def}{=} \beta /(\gamma -1)$, 
(B) $\delta = \delta_{c}$,
(C) $\delta < \delta_{c}$, and (D) $\delta = 0$
for a critical value $\delta_{c}$ of the phase transition from
uncondensation to condensation of packets, or the opposite 
transition.
Since 
the derivation is the same as in Refs. \cite{Noh05a,Noh05b} 
at $\alpha = 0$, 
except with a slight modification for a general value of $\alpha$ (and 
the corresponding $\beta \approx 1 + \alpha$), 
we briefly review it in the Appendix. 
We summarize the generalized results for $0 \leq | \alpha | \leq 1$
in Table \ref{table_kc_mk} (from that for $\alpha = 0$).
In the cases (B) and (C), 
many packets condensate at nodes with degree 
$K > K_{c}$ because of the larger exponent $\beta / \delta > \beta$.
Note that the queue length occupied by packets 
is rewritten as $m_{K_{i}}$ 
taking into account the dependence on the degree $K_{i}$.

\begin{table}[htb]
\caption{Scaling of the crossover degree $K_{c}$, the mean occupation
 number $m_{K}$ at a node with degree $K$
 and $m_{hub}$ at the hub with the maximum degree
 $K_{max}$ for the cases: (A) $\delta > \delta_{c}$, 
(B) $\delta = \delta_{c}$, (C) $0 < \delta < \delta_{c}$, 
and (D) $\delta = 0$. 
A blank denotes no correspondence.}
\begin{center}
\begin{tiny}
\begin{tabular}{l|cccc} \hline \noalign{\smallskip}
 & $K_{c}$ & $m_{K < K_{c}}$ &  $m_{K > K_{c}}$
& $m_{hub}$ \\ \noalign{\smallskip} \hline \noalign{\smallskip}
(A)
     &         &                 & $K^{\beta/\delta}$
& $O(N^{\delta_{c}/\delta})$ \\
(B)
    & $(\ln K_{max})^{\delta_{c}/\beta}$
           & $(K/K_{c})^{\beta}$ & $(K/K_{c})^{\beta/\delta_{c}}$
& $O(N/\ln N)$ \\
(C)
    & $K_{max}^{1 - \delta/\delta_{c}}$
          & $(K/K_{c})^{\beta}$ & $(K/K_{c})^{\beta/\delta}$
& $O(N)$ \\
(D) 
    & $K_{max}$ & $K^{\beta} / (K_{max}^{\beta} -  K^{\beta})$
          & & $\rho N$ \\ \noalign{\smallskip} \hline \noalign{\smallskip}
\end{tabular}
\end{tiny}
\end{center}
\label{table_kc_mk}
\end{table}

\section{Simulation} \label{sec4}
In subsection 4.1,
we numerically investigate 
the basic properties of packet transfer in the ZRP. 
In subsection 4.2, 
we further discuss congestion phenomena with 
packet generation.

\subsection{Traffic properties for $\alpha$-random walks in the 
ZRP}
We have performed simulations for $M = 1000$ packets 
on SF networks generated by the 
BA model \cite{Albert02} with a size $N = 1000$, an average degree 
$\langle K \rangle = 10$, and an exponent $\gamma = 3$
of $P(K) \sim K^{- \gamma}$. 
From the initial state in which one packet is set on each node, 
the following processes are repeated as the relaxation 
of the ZRP \cite{Noh05b} from node-based dynamics
to particle-based dynamics \cite{Evans00}.
At each time step, a packet is selected at random. 
With probability 
 $q_{i}(m_{(i)}) / m_{(i)} = m_{(i)}^{\delta - 1}$,
the packet jumps out of its resident node $i$,  
and hops to one of the neighboring nodes 
$j \in {\cal N}_{i}$ with probability 
$K_{j}^{\alpha} / \sum_{j' \in {\cal N}_{i}} K_{j'}^{\alpha}$.
Otherwise, 
the selected packet does not move 
with probability $1 - q_{i}(m_{(i)}) / m_{(i)}$. 
The time is measured as a unit of one round (Monte Carlo sweep) 
consisting of $M$ trials of the random selection of a packet.

\begin{table}[htb]
\caption{Classification of the cases in Table \ref{table_kc_mk}
 for the combinations of $\alpha$ and $\delta$.
 The values in the mid-columns are
 $K_{c} = K_{max}^{1 - \delta/\delta_{c}}$, 
 for $\delta_{c} = \beta / 2$ in the case (C), corresponding to
 the maximum, average, and minimum of $K_{max} = 189, 132, 98$
 in each vertical triplet from the top to the bottom.
 These results are obtained in 100 realizations
 of BA networks with $N = 1000$ and $\langle K \rangle = 10$.}
\begin{center}
\begin{footnotesize}
\begin{tabular}{ccc|cccc|c} \hline \noalign{\smallskip}
         & $\delta$ & & 0.0 & 0.2 & 0.5 & 0.8 & $\delta_{c}$\\
$\alpha$ &       &  &     &        &       &     &     \\
 \noalign{\smallskip} \hline \noalign{\smallskip}
         &       &  &     & 74.26  & 18.29 & 4.504 &    \\             
     1.0 &       &  & (D) & 55.29  & 14.98 & 4.063 & 1.122 \\
         &       &  &     & 43.28  & 12.71 & 3.731 & \\ 
 \noalign{\smallskip} \hline \noalign{\smallskip}
         &       &  &     & 51.44  & 7.304 & 1.037 &     \\
     0.5 &       &  & (D) & 39.27  & 6.374 & 1.034 & 0.805 \\
         &       &  &     & 31.39  & 5.693 & 1.032 &  \\ 
 \noalign{\smallskip} \hline \noalign{\smallskip}
         &       &  &     & 25.15  & 1.221  &     &     \\
     0.0 &       &  & (D) & 20.16  & 1.204  & (A) & 0.519 \\
         &       &  &     & 16.79  & 1.191  &     &     \\
 \noalign{\smallskip} \hline \noalign{\smallskip}
         &       &  &     & 3.478  &       &     & \\ 
    -0.5 &       &  & (D) & 3.193  & (A)   & (A) & 0.262 \\
         &       &  &     & 2.975  &       &     &     \\
 \noalign{\smallskip} \hline \noalign{\smallskip} 
    -1.0 &       &  & (D) & (A)    &  (A)  & (A) &  0.011 \\
 \noalign{\smallskip} \hline 
\end{tabular}
\end{footnotesize}
\end{center}
\label{table_case}
\end{table}

\begin{figure}
  \begin{minipage}[htb]{.97\hsize}
    \includegraphics[height=62mm]{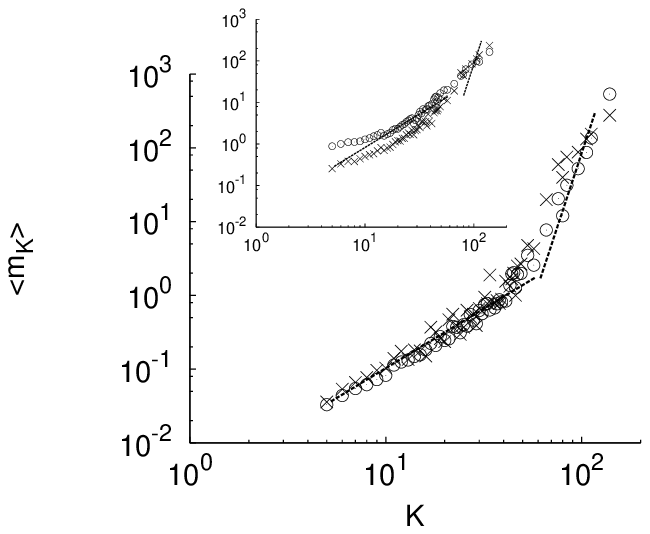}
  
    \hspace{3cm} Case(C): $\alpha =0.5$, $\delta = 0.2$
  \end{minipage}
  \hfill 

  \begin{minipage}[htb]{.97\hsize}
    \includegraphics[height=62mm]{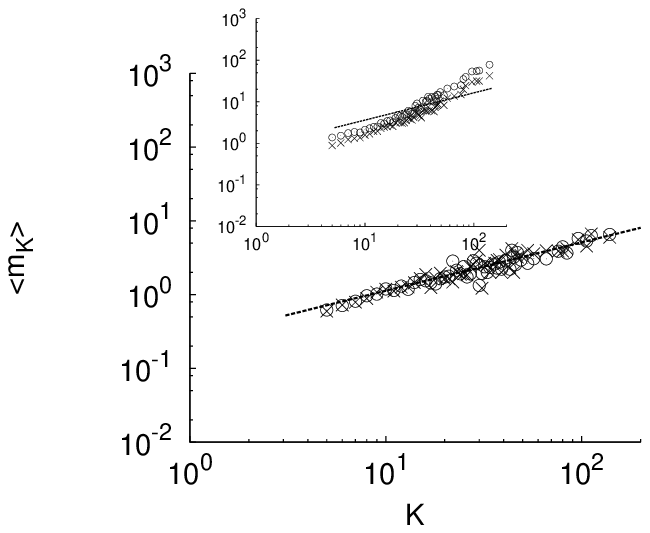}
  
    \hspace{3cm} Case(A): $\alpha =-0.5$, $\delta = 0.8$
  \end{minipage}
 \caption{Typical results for the mean occupation number
 $\langle m_{K} \rangle$
 of packets at a node with degree $K$ in the routing
 without n-search.
 Inset: the cases with n-search.
 The circle and cross marks correspond to 
 the results in 1000 and 100 rounds, respectively.
 The dashed lines show the slopes $\beta$ and $\beta / \delta$
 in Table \ref{table_kc_mk}.
 Note that condensation of packets
 at the nodes with high  degrees occurs in Case(C).
 These results are obtained from the averages of 100 samples 
 of packet transfer  
 on a BA network whose maximum degree is the closest to the 
 average value $K_{max} = 132$ in the 100 realizations.}
 \label{fig_mk}
\end{figure}

Figure \ref{fig_mk}
shows the mean occupation number $\langle m_{K} \rangle$
of packets at a node with degree $K$ 
in the cases both with/without neighbor search (n-search),
while the top of Fig. \ref{fig_mk} shows 
the predicted piecewise linear behavior \cite{Tang06} 
for the crossover degree $K_{c}$ shown in Table \ref{table_case}; 
the steeper line indicates condensation at the nodes 
with high  degrees, and 
the bottom of Fig. \ref{fig_mk} shows that 
condensation is suppressed by a more gentle line. 
As shown in the insets, 
the marks slightly deviate from a line,
especially in the head and the tail,
because of the effect of n-search into a terminal node 
at the last step.
In these cases at the top and the bottom of Fig. \ref{fig_mk}, 
the critical values of 
$\delta_{c} \approx (1+\alpha)/(\gamma - 1)$
are $0.75 > \delta$ and $0.25 < \delta$, respectively.
Thus, condensation of packets at hubs  
can be avoided as the performance of transfer is enhanced at 
a large $\delta$, 
although the transition depends on the probability of incoming packets 
according to the value of $\alpha$; 
a negative $\alpha$ induces 
a nearly homogeneous visiting of nodes, 
while a positive $\alpha$ induces a heterogeneously biased visiting of 
the nodes with high degrees.
We note that the uncondensed phase is maintained 
in a wide range of 
$\delta > \delta_{c} \approx (1+\alpha)/(\gamma - 1)$
for $\alpha < 0$, 
as shown in the case (A) of Table \ref{table_case}.

Next, in order to study the traffic properties, such as the travel 
time on the routing path, 
we consider packet dynamics 
in a realistic situation with 
n-search into a terminal node at the last step. 
If the terminal is included among the neighbors of the 
resident node for a randomly selected packet, 
then it is deterministically forwarded to the terminal node,
taking into account the reachable chance, 
otherwise it is stochastically forwarded to a neighboring node $j$
with probability 
$K_{j}^{\alpha} / \sum_{j' \in {\cal N}_{i}} K_{j'}^{\alpha}$.
The n-search 
is practically effective and necessary in order to reach a terminal node.
Remember that 
the estimated values of $\beta$ are similar to both 
with/without n-search, 
as shown in Table \ref{table_est_beta}.
Therefore, the behavior of 
the mean occupation number $\langle m_{K} \rangle$ 
is similar to that shown in Fig. \ref{fig_mk} and the inset.
In the following discussions, 
if we leave out n-search, then 
the optimal parameter values of $\alpha$ and $\delta$ 
may be changed for low-latency delivery, 
although the difference is probably small from the above similarity.
By selecting hubs for $\alpha > 0$, a packet is terminated with higher
probability even in only the stochastic forwarding, 
because a terminal node is highly likely to be connected to some hubs.
At the same time, this leads to congestion at hubs.
However, in the case without n-search, 
a packet may wander for a very long time, which is not bounded 
a priori for the simulation of the packet transfer.
It is intractable  due to huge computations.
Thus, we focus on the case with n-search.

We investigate the traffic properties for reachability of
packets, the number of hops, the travel time $T_{a}$ 
including the wait time of a trapped packet in queues, 
and the sum $T_{w}$ of wait times 
on a path until arrival at the terminal. 
Note that the travel time is 
$T_{a} = T_{w} \;\; +$ 
(time to one hop)$\times$(the number of hops).
We also define the averaged wait time $T_{w} / N_{w}$ per node, 
where $N_{w}$ is the number of trappings in queues at the nodes 
on a routing path.
These measures 
are cumulatively counted in the observed interval after 1000 rounds,
and averaged over 100 samples of this simulation. 
Here, we discarded the initial 1000 rounds as a transient before 
the stationary state of $m_{(i)}$ is reached.

\begin{figure}
  \begin{minipage}[htb]{.47\textwidth}
    \includegraphics[height=50mm]{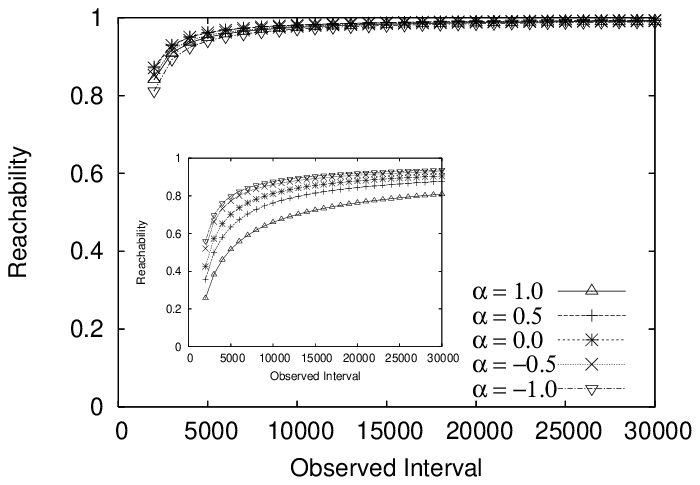}
  \end{minipage}
  \hfill

  \begin{minipage}[htb]{.47\textwidth}
    \includegraphics[height=50mm]{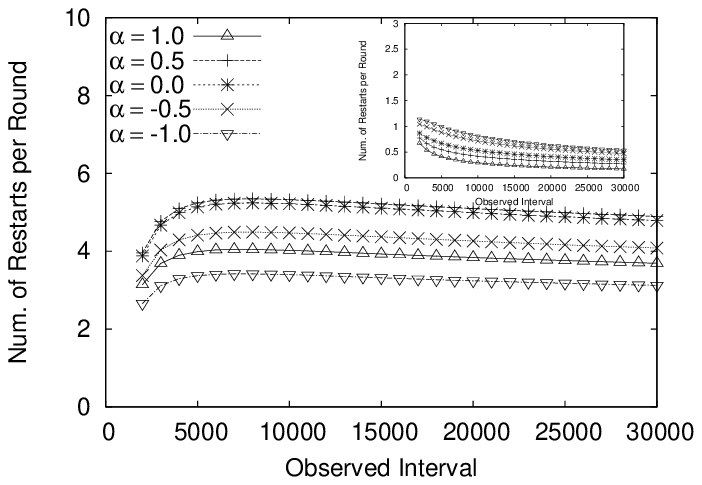}
    \end{minipage}
  \hfill

  \begin{minipage}[htb]{.47\textwidth}
    \includegraphics[height=50mm]{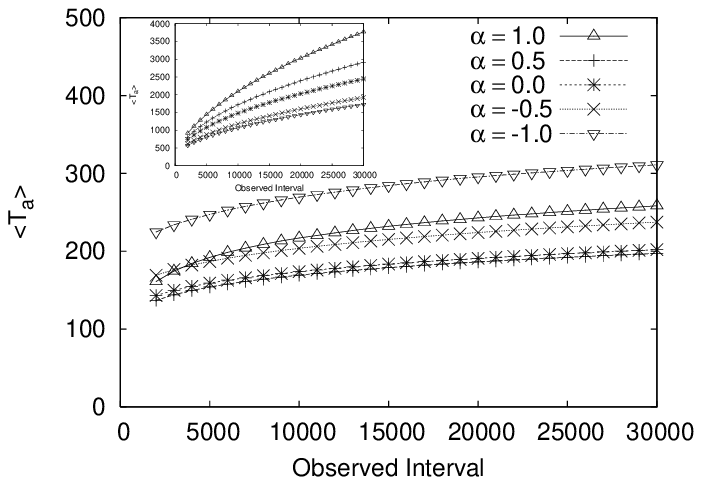}
  \end{minipage}
\caption{Convergence properties 
 in  the high-performance regime at $\delta = 0.8$
 in the observed interval [rounds]
 after the transient state of 
 $\langle m_{K} \rangle$.
 Inset: results in  the low-performance regime at $\delta = 0.0$. 
 These results are obtained from the averages of 100 samples 
 of packet transfers 
 on a BA network whose maximum degree is the closest to the 
 average value $K_{max} = 132$ in the 100 realizations.}
 \label{fig_reach}
\end{figure}

\begin{figure}
  \begin{minipage}[htb]{.47\textwidth}
    \includegraphics[height=45mm]{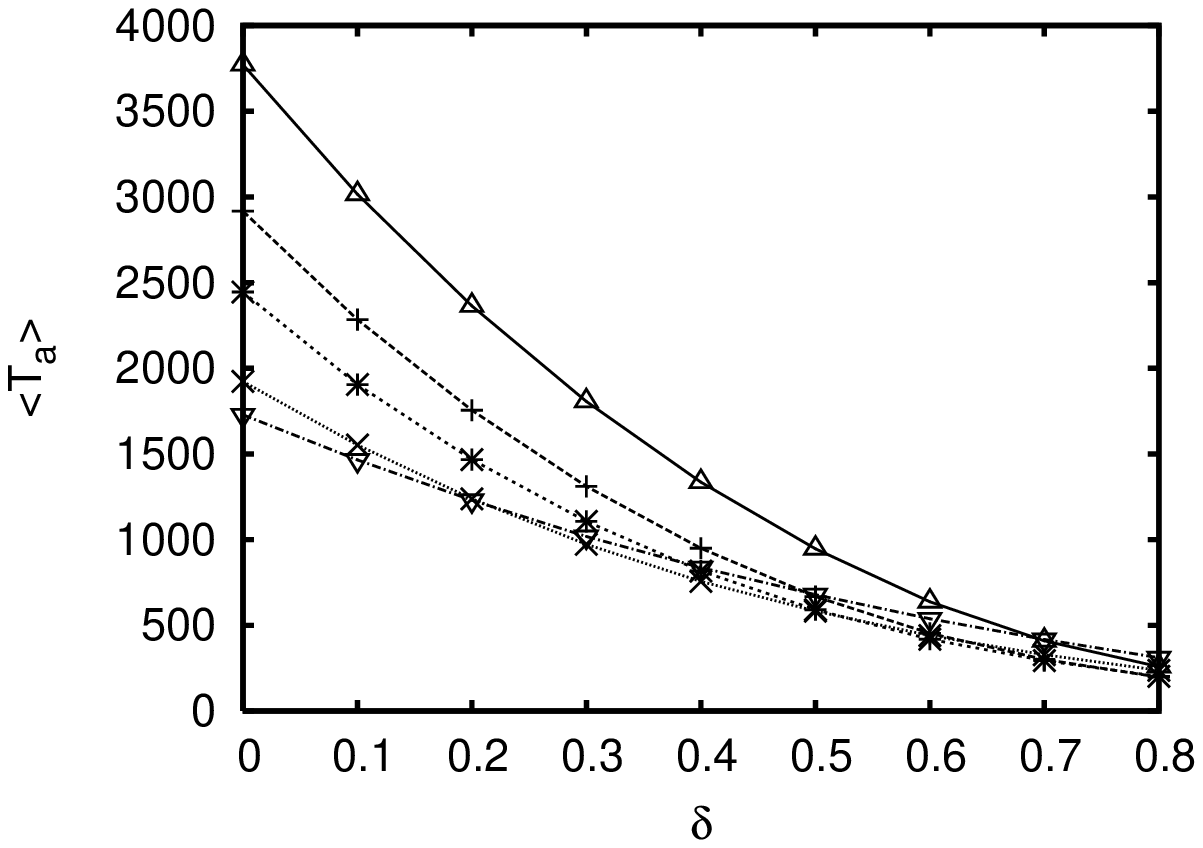}
  \end{minipage}
  \hfill

  \begin{minipage}[htb]{.47\textwidth}
    \includegraphics[height=45mm]{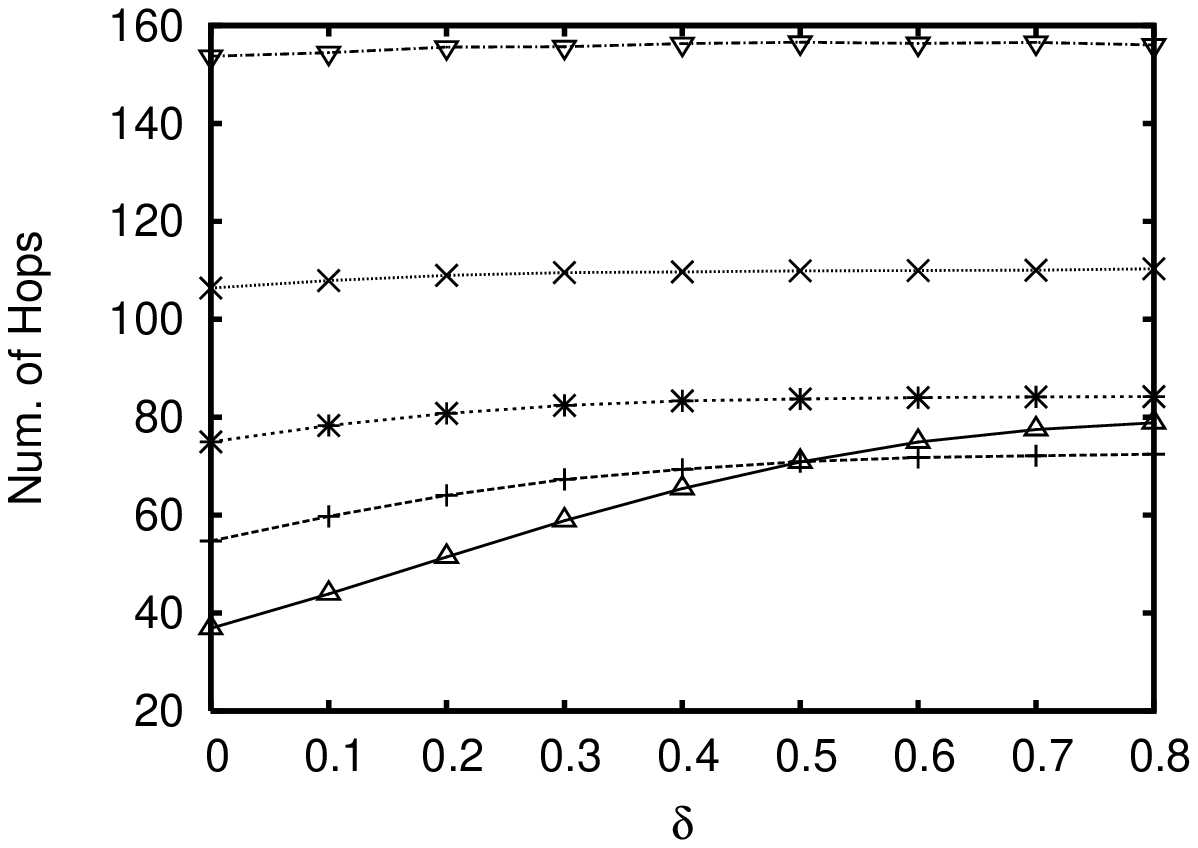}
  \end{minipage}
  \hfill
 \caption{The mean travel time (top) and 
 the mean number of hops (bottom).
 The $\bigtriangleup$, $+$, $\ast$, $\times$, 
 and $\bigtriangledown$ marks
 correspond to $\alpha = 1.0, 0.5, 0.0, -0.5$, and $-1.0$, respectively.
 The simulation 
 conditions are the same as in Fig. \ref{fig_reach}.}
 \label{fig_traffic_prop1}
\end{figure}

Figure \ref{fig_reach} shows, from the top to the bottom, 
the convergence of reachability, 
the mean number of restarted packets per round, 
and the mean travel time 
in the high-performance regime at $\delta = 0.8$.
The inset shows a slightly slow convergence 
in the low-performance regime at $\delta = 0.0$.
For other measures, 
similar convergence properties are obtained.
We compare these curves in terms of 
the values of $\alpha$; they shift up from $\alpha = -1.0$
to $\alpha = 0.5$ in the case (A) $\delta > \delta_{c}$, 
while down from $\alpha = 0.5$ to $\alpha = 1.0$
in the case (C) $\delta < \delta_{c}$.
This non-monotonic dependence on $\alpha$ is related to the condensation
transition, since the change between $\delta > \delta_{c}$ and 
$\delta < \delta_{c}$ occurs at the critical value 
$\alpha = 0.6$ for the equivalence 
$\delta_{c} \approx (1+\alpha)/(\gamma - 1)$ $\Leftrightarrow$
$\delta = 0.8$. 
Thus, it appears when $\alpha$ is greater or less than $0.5$ 
in Fig. \ref{fig_reach}.
In the following, we briefly explain each property: 
reachability, the number of
restarts, and $\langle T_{a} \rangle$. 
The reachability of the restarted packets is around 
$0.99$ on similar curves for all values of $\alpha$ 
at $\delta = 0.8$, 
while these curves separate 
in increasing order of $\alpha$ at  $\delta = 0.0$
(see the inset at the top of Fig. \ref{fig_reach}). 
Note that the number of restarted packets cumulatively increases 
as the observed interval is longer, although the rate per round is 
constantly around $3 \sim 5$ in the ordering 
from $\alpha = 0$, $\alpha = \pm 0.5$, 
to $\alpha = \pm 1$ for $\delta = 0.8$, 
and less than $1$ in the ordering from $\alpha = -1.0$
to $\alpha = 1.0$ for $\delta = 0.0$, 
as shown in the middle of Fig. \ref{fig_reach} and in the inset.
The maximum and the minimum lines for 
the mean travel time $\langle T_{a} \rangle$ 
are obtained at 
$\alpha = -1$ and $\alpha = 0$, respectively, for $\delta = 0.8$.
However, all of the curves shift up for $\delta = 0.0$
(see the inset at the bottom of Fig. \ref{fig_reach}),
and $\langle T_{a} \rangle$ is longer 
as the value of $\alpha$ increases, 
because of the waiting at high-degree nodes.

We further investigate the above traffic properties, especially 
for the forwarding of packets with 
more detailed values of $\delta$ for 
30000 rounds in the quasi-convergent state. 
As shown in Fig. \ref{fig_traffic_prop1}, 
the mean travel time $\langle T_{a} \rangle$ decreases 
as the value of $\delta$ increases with higher forwarding 
performance, because the wait time trapped in a queue 
decreases on average.   
In particular, packets tend to be trapped at hubs 
for a long time when $\alpha > 0$, and then 
$\langle T_{a} \rangle$ is longer. 
In contrast, 
the number of hops increases on average
as the value of $\delta$ increases, because some 
longer paths are included in higher reachability. 
The path length counted by hops 
tends to be short through hubs when 
$\alpha > 0$, 
however it tends to be long on a wandering path when 
$\alpha < 0$.
Therefore the number of hops 
increases in decreasing order of $\alpha$. 
Note that the number of hops is very small compared to 
$\langle T_{a} \rangle$, which is dominated by 
the mean wait time 
$\langle T_{w} \rangle \approx \langle T_{a} \rangle$ 
on a path.

\begin{figure}
  \begin{minipage}[htb]{.47\textwidth}
    \includegraphics[height=50mm]{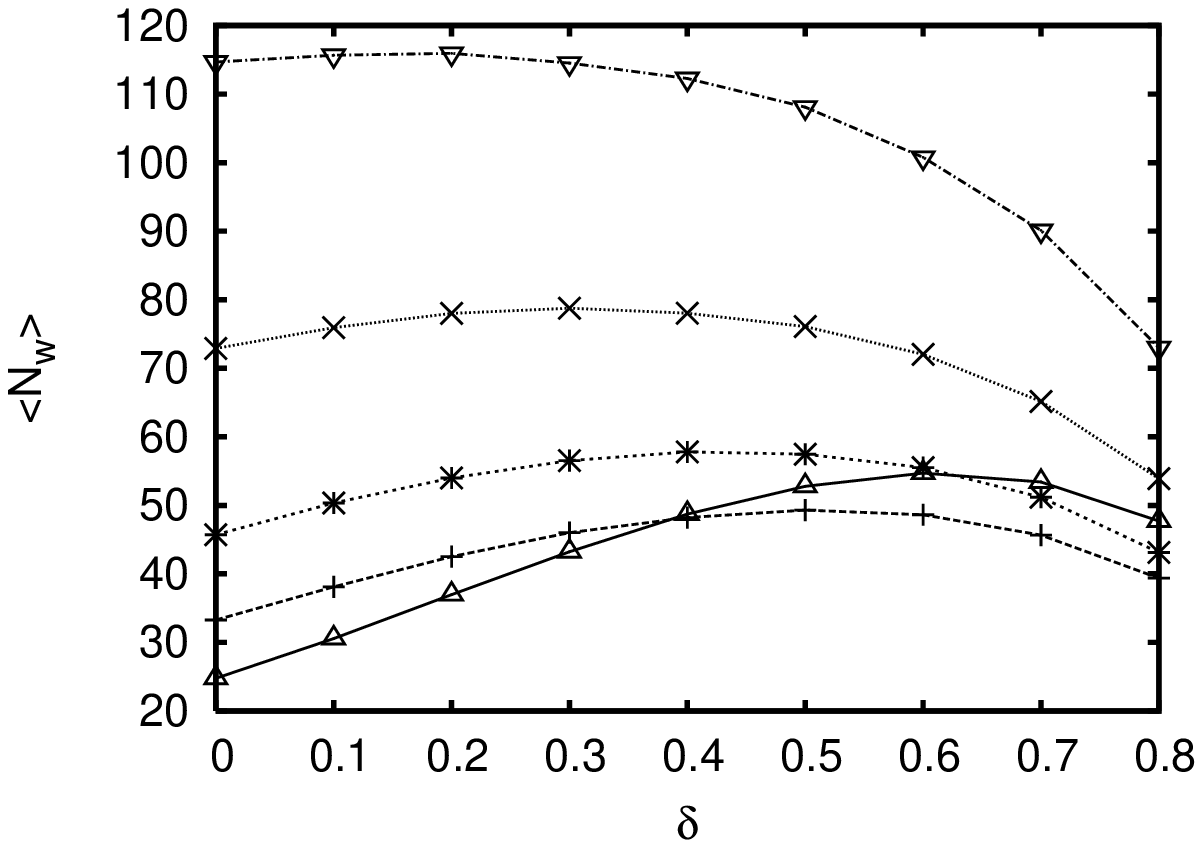}
  \end{minipage}
  \hfill

  \begin{minipage}[htb]{.47\textwidth}
    \includegraphics[height=50mm]{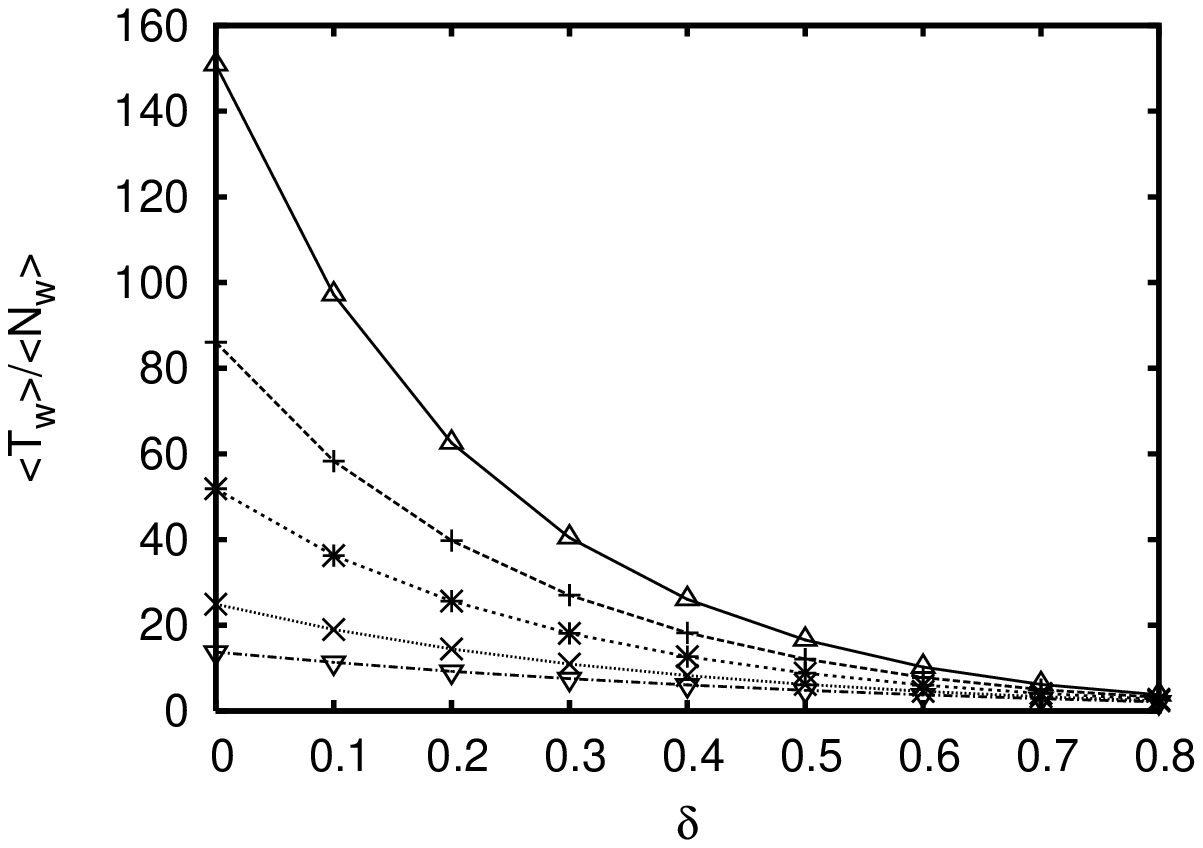}
  \end{minipage}
 \caption{The mean number of trappings at nodes (top) 
 and the mean wait time per node (bottom).
 The $\bigtriangleup$, $+$, $\ast$, $\times$, 
 and $\bigtriangledown$ marks 
 correspond to $\alpha = 1.0, 0.5, 0.0, -0.5$, and $-1.0$, respectively. 
 The simulation 
 conditions are the same as in Fig. \ref{fig_reach}.}
 \label{fig_traffic_prop2}
\end{figure}

As shown in Fig. \ref{fig_traffic_prop2}, 
the mean number 
$\langle N_{w} \rangle$ of trappings at nodes on a path 
increases when $\alpha > 0$
and decreases when $\alpha < 0$
as the value of $\delta$ increases with higher 
forwarding performance. 
This up-down phenomenon 
may be caused by a trade-off between the avoidance of 
trapping through sufficiently 
high forwarding performance at a node 
and the inclusion of longer paths with high reachability. 
The mean wait time $\langle T_{w} / N_{w} \rangle$ per node 
decreases as the value of $\delta$ increases, 
and is longer in increasing order of $\alpha$
because of the long wait time at hubs.

In summary, 
the wandering path for $\alpha < 0$ better reduces
the mean travel time
of a packet with high reachability 
in the low-performance regime at a small $\delta$,
while in the high-performance regime at a large $\delta$,
the difference between $\alpha > 0$ and $\alpha < 0$ is small,
neither the
wandering long path with short wait trapped at nodes
($\alpha = -1$),
nor the short hopping path
with long wait trapped at hubs ($\alpha = 1$) is advisable.
Thus, a uniformly random walk ($\alpha = 0$)
yields slightly better performance.

\subsection{Congestion phenomenon}
This subsection discusses the congestion phenomenon 
when packets are randomly generated at each node at a constant rate 
$p$, and removed at the terminal nodes 
(not restarted within the persistency).
In order to reduce the computational load, 
the packet dynamics starts from the initial state: 
there are no other packets than the ones that are generated.
We compare the phenomenon
in our traffic model based on the ZRP 
with that in the following modifications 
related to the traffic-aware routing \cite{Echenique05} 
at $\alpha = 0$ \cite{Martino09}.
Table \ref{table_modified_models} summarizes a combination of the basic
processes: with or without (Yes or No) the refusal of forwarding,
n-search, and a constant arrival with probability $\mu$.
The other processes for choosing a forwarding node $j$ 
with probability
$\propto K_{j}^{\alpha}$ 
and for jumping-out a packet from its resident node $i$ 
at the rate $m_{(i)}^{\delta}$ 
are common.

\begin{description}
  \item[Mod 1:] With probability 
  $\eta(m_{(j)}) = \bar{\eta} \Theta(m_{(j)} - m^{*})$,
  the selected packet is not transfered 
  to $j \in {\cal N}_{i}$, 
  but is refused at node $i$, 
  where $\Theta(x)$ is the step function, 
  $m^{*}$ is a threshold, and a parameter $0 < \bar{\eta} \leq 1$.\\
  Here, node $j$ is chosen 
  deterministically by n-search if $j$ is the terminal node,
  otherwise it is chosen with a probability 
  proportional   to $K_{j}^{\alpha}$.
  \item[Mod 2:] Moreover, instead of n-search,
  the selected packet is removed with probability $\mu$,
  or 
  it enters the queue with probability $1 - \mu$. 
  \item[Mod 3:] In Mod 2                                 , 
	     there is no refusal process ($\bar{\eta} = 0$). 
\end{description}
In Mod 1, 
the randomly-selected packet from the queue does not leave node $i$ 
with a constant probability $\bar{\eta}$ 
if the occupation number $m_{(j)}$ of packets 
is greater than a threshold $m^{*}$, 
while in Mod 2, 
after passing this refusal check, 
it is removed with a constant probability $\mu$. 
Note that constant arrival was assumed to 
theoretically predict the critical point of traffic 
congestion in the mean-field approximation as 
$N \rightarrow \infty$ \cite{Martino09}. 
With packet generation, 
we can perform the ZRP as an extension of the model 
in Ref. \cite{Martino09}.
In particular, 
the case of $\delta = 0$ corresponds to 
 node capacity $c_{i} = 1$ for all nodes $i$;
only one packet is transferable from a node at each time.

\begin{table}[htb]
\caption{Modified traffic models.}
\begin{center}
\begin{tabular}{c|c|c} \hline \noalign{\smallskip}
Refusal & n-search & const. arrival 
\\  \noalign{\smallskip} \hline \noalign{\smallskip}
No $\bar{\eta} = 0$ & ZRP   & Mod 3 \\
Yes $\bar{\eta} > 0$ & Mod 1 & Mod 2 \\
\noalign{\smallskip} \hline 
\end{tabular}
\end{center}
\label{table_modified_models}
\end{table}

For a variable generation rate $p$, 
we investigate the appearance of congestion 
by using the order parameter \cite{Echenique05,Martino09}
\[
  op = \lim_{t \rightarrow \infty} \frac{M(t + \tau) - M(t)}{\tau p N},
\]
where $M(t)$ denotes the sum of existing packets in queues over 
network (practically for a large $t$), and 
$\tau$ is the observed interval.
The value of $op$ represents level of congestion, e.g. 
$op \approx 0$ indicates a free-flow regime, while $op \approx 1$ 
indicates a congested regime.

\begin{figure}
  \begin{minipage}[htb]{.47\textwidth}
    \includegraphics[height=60mm,angle=-90]{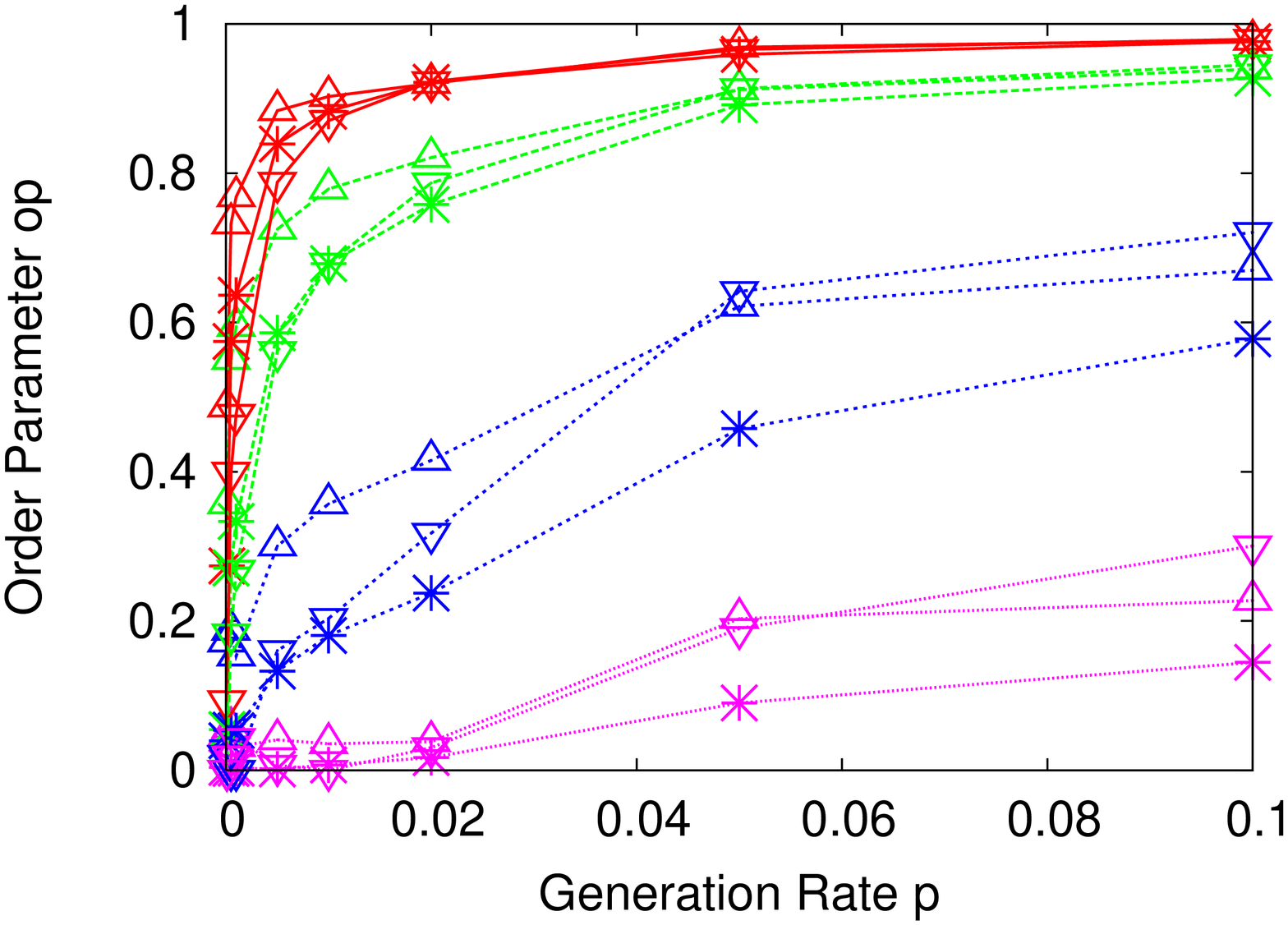}
  \end{minipage}
  \hfill

  \begin{minipage}[htb]{.47\textwidth}
    \includegraphics[height=60mm,angle=-90]{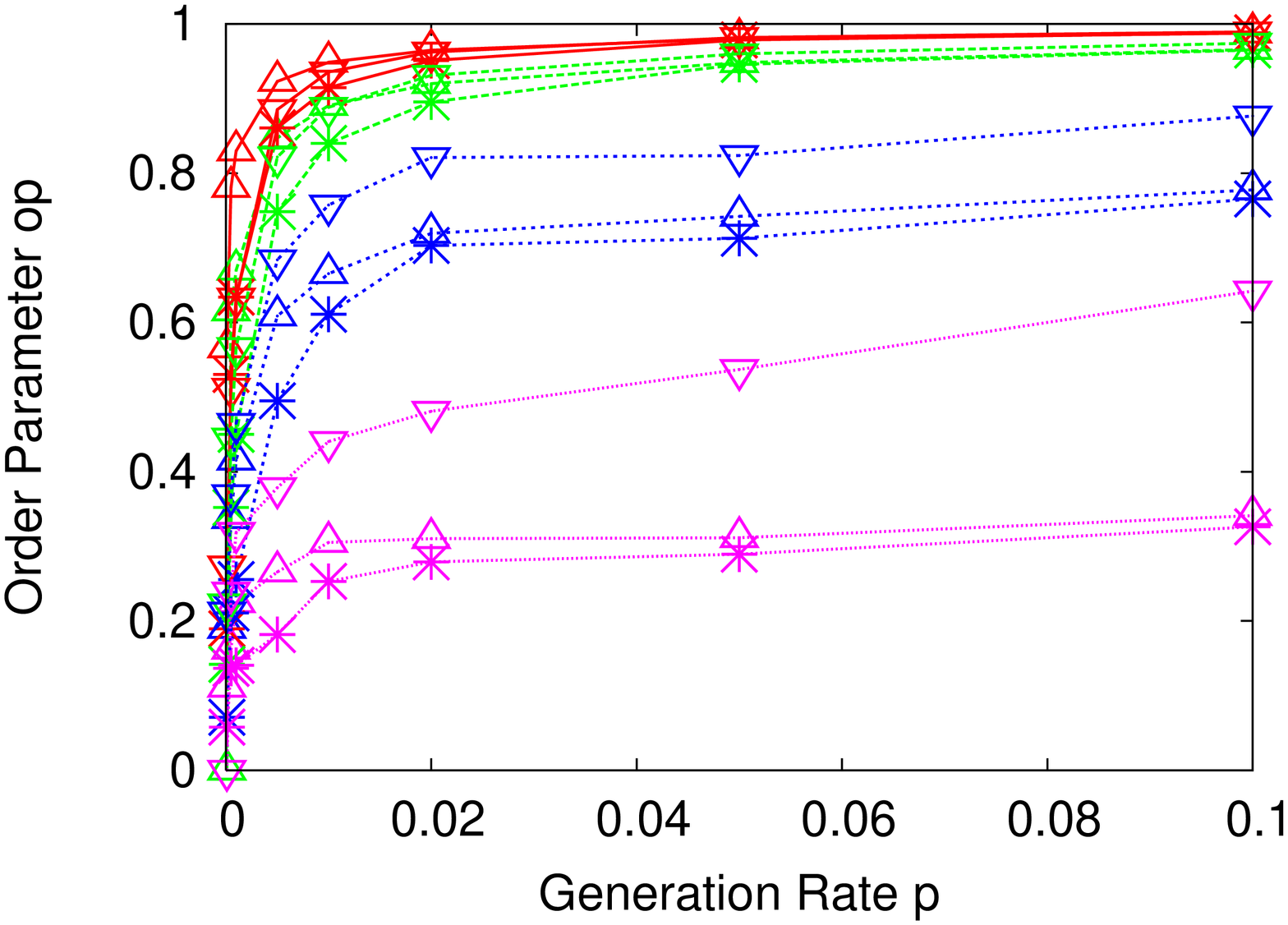}
  \end{minipage}
 \caption{The values of 
 $op$ for the ZRP (top) and Mod 1 (bottom) using n-search.
 The open $\bigtriangleup$, $\ast$, 
 and $\bigtriangledown$ marks 
 correspond to $\alpha = 1.0, 0.0$, and $-1.0$, respectively. 
 The red, green, blue, and magenta lines correspond to 
 $\delta= 0.0, 0.2, 0.5$, and $0.8$, respectively.}
 \label{fig_congestion_zrp}
\end{figure}

In the following, 
we set $\bar{\eta} = 0.7$ and $m^{*} = 5$
for the refusal process.
As shown in Fig. \ref{fig_congestion_zrp}, 
in the cases in which n-search takes place, 
the value of $op$ rapidly grows 
with the increasing of the generation rate $p$, 
since the removal of a packet arriving at the terminal node 
is rare, especially for a small $\delta$. 
Here, 
the marks and color lines indicates different values 
of $\delta$ and $\alpha$.
By comparing the corresponding curves 
of the same color and marks in the top and the bottom figures, 
we notice that 
the ZRP suppresses congestion in   smaller $op$s 
than Mod 1, in particular, for $\delta = 0.5, 0.8$
(blue and magenta lines).
At the top of Fig. \ref{fig_congestion_zrp}, 
the magenta lines indicate 
the existence of a free-flow regime 
around a small $p$ for $\delta = 0.8$ as 
high forwarding performance.
Thus, the refusal process does not work effectively 
in the cases with n-search, 
at least for this parameter set. 
At both the top and the bottom of Fig. \ref{fig_congestion_zrp}, 
the difference for the 
same color lines with three marks corresponding to 
$\alpha = \pm 1, 0$ is small, 
except for $\delta = 0.8$ in Mod 1 (magenta lines at the bottom).
However, the curves shift down 
as $\delta$ becomes larger; in other words, 
the congestion is suppressed by higher forwarding performance. 
Note that a uniformly random walk 
at $\alpha = 0$ (asterisk marks) 
yields better performance for each value of $\delta$.

\begin{figure} 
  \begin{minipage}[htb]{.47\textwidth}
    \includegraphics[height=60mm,angle=-90]{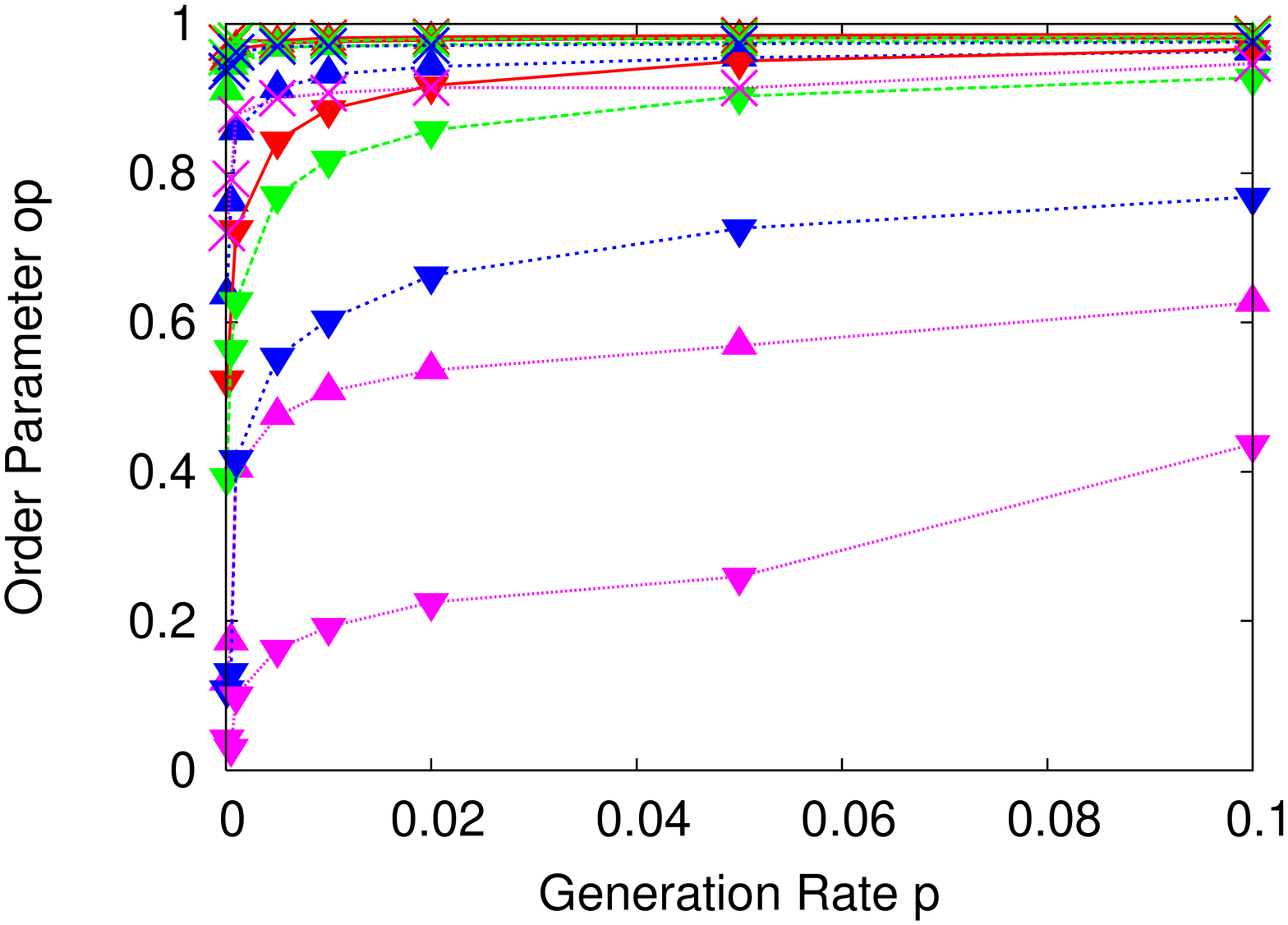}
  \end{minipage}
  \hfill

  \begin{minipage}[htb]{.47\textwidth}
    \includegraphics[height=60mm,angle=-90]{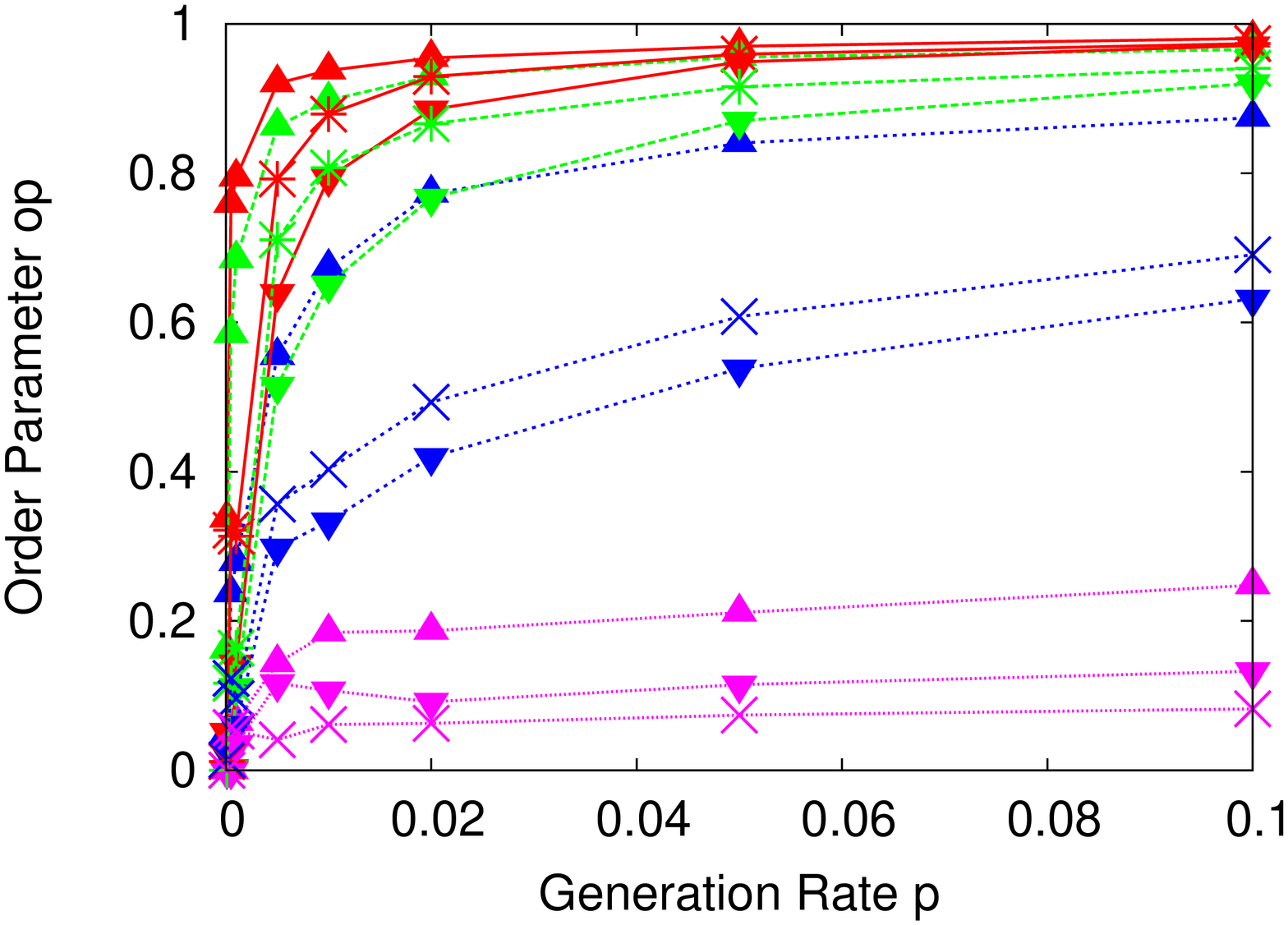} 
  \end{minipage}
 \caption{The values of 
 $op$ for Mod 3 (top) and Mod 2 (bottom) 
 at a constant arrival with probability $\mu = 0.01$,
 instead of using n-search.
 The filled $\bigtriangleup$, $\times$, 
 and $\bigtriangledown$ marks 
 correspond to $\alpha = 1.0, 0.0$, and $-1.0$, respectively.
 The red, green, blue, and magenta lines correspond to 
 $\delta= 0.0, 0.2, 0.5$, and $0.8$, respectively.}
 \label{fig_congestion_reject}
\end{figure}

When a constant arrival with probability $\mu$ is applied 
instead of the realistic n-search, 
the behavior changes.
Figure \ref{fig_congestion_reject} shows that 
the refusal process works effectively, 
since 
Mod 2 with the refusal process (at the bottom)
has smaller $op$ 
than Mod 3 without the refusal process (at the top).
The gap between three marks for the lines of each color 
for Mod 2 at the bottom of Fig. \ref{fig_congestion_reject}
resembles to that at the top of Fig. \ref{fig_congestion_zrp}, 
however the gap appears remarkably 
at $\delta = 0.8$ (magenta lines) for Mod 3
in the top of Fig. \ref{fig_congestion_reject}. 
Although n-search leads to a low arrival rate 
(lower than $\mu = 0.01$) as shown in Fig. \ref{fig_reach},
and reachability is not $100 \%$ 
even in the case of persistent packets after restarting,  
in the meaning of smaller $op$,
the ZRP is better than other models, 
by comparison with the corresponding curves  
in Figs. \ref{fig_congestion_zrp} and \ref{fig_congestion_reject}.

\begin{figure}
 \begin{center}
   \includegraphics[height=60mm,angle=-90]{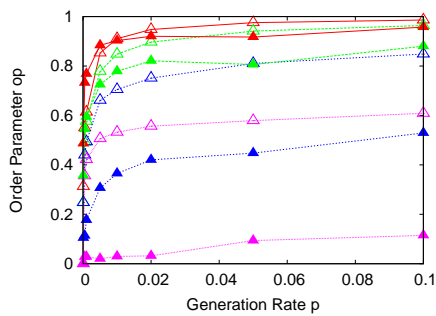}
 \end{center}
 \caption{The values of 
 $op$ for a traffic-aware routing \cite{Wang06b} 
 in which a forwarding node $k$ 
 is chosen by applying Eq.(\ref{eq_km-walk}) in the cases of 
 n-search (open marks), 
 and of a constant arrival with probability $\mu = 0.01$ (filled marks).
 The red, green, blue, and magenta lines correspond to 
 $\delta= 0.0, 0.2, 0.5$, and $0.8$, respectively.}
 \label{fig_congestion_km}
\end{figure}

We consider the other stochastic routing method in which 
a forwarding node $k \in {\cal N}_{i}$ is chosen with probability 
\begin{equation}
  \Pi_{k} \propto K_{k}(m_{(k)}+1)^{-\beta},
  \label{eq_km-walk}
\end{equation}
where $\beta = 3$ yields 
the maximum generation rate in a free-flow
regime \cite{Wang06b}. 
As shown in Fig. \ref{fig_congestion_km}, 
in this optimal case, 
the behavior is similar to that in Mod 3 without the refusal process 
at the top of Fig. \ref{fig_congestion_reject}, 
although it has better performance than Mod 3. 
There is a free-flow regime in the case when $\delta = 0.8$ 
(the magenta line with filled upward-pointing triangle marks)
at a constant arrival.
In this method, 
a forwarding node is dynamically  selected in a balance between 
reducing distance by passing through large degree nodes, 
and avoiding congestion.
Thus, a further improvement may be potentially expected
in the tuning of the balance for $\alpha$-random walks
and other modifications.

\section{Conclusion} \label{sec5}
For a SF network, whose topology is found in many real systems, 
we have studied extensions of 
the ZRP \cite{Noh05a,Noh05b,Noh07,Noh04} 
which controls both the routing strategies in the 
preferential \cite{Yamashita03,Wang06a,Yan06}
and congestion-aware \cite{Danila06} walks, 
and the node performance for packet transfers.
Under the assumption of persistent packets, 
we have approximately analyzed the phase transition between 
condensation of packets at hubs and uncondensation on SF networks 
by another straightforward approach \cite{Noh05a,Noh05b} 
instead of the mean-field approximation in the preferential walk for  
$\alpha >0$ \cite{Tang06}. 
In particular, 
we have found that uncondensation is maintained in a
wide rage of 
$\delta > \delta_{c} \approx  (1+\alpha)/(\gamma - 1)$ 
for $\alpha < 0$.

\begin{table}[htb]
\caption{Qualitative summary of the traffic properties.}
\begin{center}
\begin{scriptsize}
\begin{tabular}{c|cc|c} \hline \noalign{\smallskip}
Mean Value
 & $\alpha > 0$ & $\alpha < 0$ & for larger $\delta$ \\  \noalign{\smallskip} \hline \noalign{\smallskip}
Reachability & low   & {\bf high}  & increased $\nearrow$ \\
Num. of Hops & {\bf small} & large & increased $\nearrow$ \\
Travel Time $\langle T_{a} \rangle$
             & long  & {\bf short} & decreased $\searrow$ \\
One Wait $\langle T_{w} \rangle / \langle N_{w} \rangle$
             & long  & {\bf short} &  decreased $\searrow$ \\
Num. of Wait $\langle N_{w} \rangle$
             & {\bf small} &       & increased $\nearrow$ \\
             &       & large & decreased $\searrow$ \\ \noalign{\smallskip} \hline \noalign{\smallskip}
Characteristics
             & condensation & uniformly \\
of a Path    & at hubs   & wandering \\ \noalign{\smallskip} \hline
\end{tabular}
\end{scriptsize}
\end{center}
\label{table_traffic_summary}
\end{table}

Moreover, we have numerically investigated the traffic properties 
when the practical n-search into a terminal node 
at the last step takes place.
The phase transition has been consistently observed in the cases both 
with/without n-search.
The obtained results are summarized 
in Table \ref{table_traffic_summary}. 
We conclude that 
the wandering path for $\alpha < 0$ better reduces
mean travel time
of a packet with high reachability 
in the low-performance regime at a small $\delta$,
while in the high-performance regime at a large $\delta$,
neither the
wandering long path with short wait trapped at nodes
($\alpha = -1$),
nor the short hopping path
with long wait trapped at hubs ($\alpha = 1$) is advisable. 
A uniformly random walk ($\alpha = 0$) yields 
slightly better performance. 
This optimality at $\alpha = 0$ in a high-performance regime 
is consistent with the results obtained for the
critical generation rate in other traffic model \cite{Wang06a}
on a SF network with
node capacity (corresponding to the forwarding performance) 
proportional to its degree. 
However, in the details for our traffic model, 
the optimal case depends on a combination of the values of 
$\alpha$ and $\delta$ related to the 
condensation transition at $\delta_{c}$.
We emphasize that, for high reachability, small number of hops, and
short travel or wait time, 
such traffic properties 
summarized in Table \ref{table_traffic_summary}
as the trade-off between a detour wandering path and long wait at hubs
cannot be obtained from only the theoretically predicted 
phase transition between condensation 
and uncondensation in Ref. \cite{Noh05a,Noh05b,Noh07,Tang06}. 
Concerning the fundamental traffic properties,
we have investigated the congestion phenomenon  with 
packet generation, 
compared with other models related to the 
traffic-aware routing \cite{Wang06b,Echenique05,Martino09}.
We suggest that 
the high-forwarding performance at a large $\delta$ 
is more dominant than the refusal process
in order to suppress congestion in a small $op$, 
and that the difference of $op$ is small when 
the values of $\alpha$ are varied 
(the case of $\alpha = 0$ is slightly better
for n-search).
The above results only show qualitative tendencies.
In further research, more complex and 
quantitative properties should be carefully 
discussed for many combinations of parameters $\alpha$, $\delta$, 
$\mu$, $\bar{\eta}$, $m^{*}$, etc., 
although such simulations may be intractable due to 
huge computation load and memory consumption
for the processing of millions of packets 
in very long iterations.

This direction of research on stochastic routing 
may be the first step to reveal 
 complex traffic dynamics such as the 
trade-off between the selection of a short path 
and long wait (delay) at some particular nodes related to 
the underlying network structure.
We will investigate this problem, including the effects of 
more realistic selections of the source and the terminal nodes
which depend on geographical positions or population density, 
the queue discipline such as FIFO or LIFO, 
and other topologies to develop the optimal routing schemes 
\cite{Tadic07} for advanced sensor or ad hoc networks.

\section*{Acknowledgment}
The authors would like to thank anonymous meta-reviewer and 
reviewers for their valuable comments. 
This research is supported in part by a 
Grant-in-Aid for Scientific Research in Japan, No. 21500072.

\appendix*
\section{}
As a modification to Refs. \cite{Noh05a,Noh05b},
we briefly review the derivation of the results presented 
in Table \ref{table_kc_mk}. 

For $\delta = 0$: case (D), 
we have a constant jumping rate 
$q_{i}(\omega) = 1$ independent of the occupation number $\omega$ 
and 
$F_{i}(z) = \sum_{\omega} \left( z K_{i}^{\beta} \right)^{\omega} 
  = \frac{1}{1 -  z K_{i}^{\beta}}$. 
Then, from Eq. (\ref{eq_mk_calc}),
\[
 m_{K_{i}} = \frac{z K_{i}^{\beta}}{1 - z K_{i}^{\beta}} 
 \rightarrow \frac{K_{i}^{\beta}}{K_{max}^{\beta} - K_{i}^{\beta}},  
\]
is divergent for the hub with the maximum degree $K_{max}$ 
as 
$z$ approaches $z_{c} = 1 / K_{max}^{\beta}$
in the range $z < z_{c}$, 
because 
the denominator becomes zero in the first term of the right-hand
side, equivalently to the case in which 
node $i$ has the maximum degree in the second term.
It is convenient to decompose 
the density into two terms such that $\rho = \rho_{s} + \rho_{n}$
for the hub with the maximum degree and for the other nodes. 
These two terms are given by 
\begin{equation}
  \rho_{s} = \frac{m_{hub}}{N} = 
  \frac{1}{N} \frac{z K_{max}^{\beta}}{1 - z K_{max}^{\beta}},
  \label{eq_rho_s_AP}
\end{equation}
\begin{equation}
  \rho_{n} = \int_{K_{min}}^{K_{max}-1} 
  \frac{z K^{\beta}}{1 - z K^{\beta}} P(K) dK.
  \label{eq_rho_n_AP}
\end{equation}
By using the relation 
$\frac{1}{1 - x} = \sum_{l = 0}^{\infty} x^{l}$ 
for Eq. (\ref{eq_rho_n_AP}), 
we derive 
\[
  \begin{array}{ll}
  \rho_{n} & = \int \frac{z K^{\beta}}{1 - z K^{\beta}} P(K) dK \\
   & = \int z K^{\beta} 
       \sum_{l=0}^{\infty} (z K^{\beta})^{l} P(K) dK \\
   & \sim  \sum_{l=1}^{\infty} \left\{ 
             \int (z K^{\beta})^{l} K^{-\gamma} dK \right\} \\
   & = \sum_{l=1}^{\infty} \left\{ z^{l}
             \int K^{\beta l - \gamma} dK \right\} \\
   & = \sum_{l=1}^{\infty} \left\{ z^{l} 
     \frac{(K_{max} - 1)^{(\beta l - \gamma + 1)}}{\beta l - \gamma + 1} 
     \right\} \\
   & \sim \sum_{l=1}^{\infty} \left\{ 
     \frac{( z K_{max}^{\beta})^{l}
     \times (K_{max}^{(-\gamma + 1)} )}{\beta l} \right\} \\
   & = \frac{K_{max}^{(-\gamma + 1)}}{\beta} 
     \times \sum_{l=1}^{\infty} \frac{(z K_{max}^{\beta})^{l}}{l} \\
   & = \frac{K_{max}^{(-\gamma + 1)}}{\beta} 
     \times [ - \ln (1 - z K_{max}^{\beta})] \\
   & \sim N^{-1} \times \ln(1 + m_{hub}) \rightarrow 0, 
  \end{array}
\]
for a large size $N \rightarrow \infty$. 
In the last terms, we apply $K_{max}^{(-\gamma + 1)} \sim N^{-1}$
from $\int_{K_{max}}^{\infty} P(K) dK \sim 1/N$, and 
\[
   \ln\frac{1}{1-x} = - \ln (1-x) 
   = \sum_{l = 1}^{\infty} \frac{x^{l}}{l},
\]
\[
   \frac{1}{1 - z K_{max}^{\beta}} = 1 + \rho_{s} N = 1 + m_{hub} > 0,
\]
from Eq. (\ref{eq_rho_s_AP}). 
Since only $\rho_{s}$ remains,
condensation occurs at the hub for any $\alpha$
in case (D) at this uniform jumping rate given by $\delta = 0$. 

In the general case where $\delta > 0$, 
the infinite series in Eq. (\ref{eq_Fi_series}) 
does not have a closed form, except for 
$F_{i}(z) = e^{z K_{i}^{\beta}}$
at $\delta = 1$. 
We approximate the series 
as in \cite{Noh05a,Noh05b}, 
\[
  F_{i}(z) \approx \frac{1}{\sqrt{\delta}} 
  (2 \pi (z K_{i}^{\beta})^{1/\delta})^{(1 - \delta)/2} \times 
  \exp( \delta (z K_{i}^{\beta})^{1/\delta}).
\]
The second term in the right-hand side 
is dominant for $z K_{i}^{\beta} \geq 1$. 
We can simply approximate it with a few lowest-order terms such as 
$F_{i}(z) = 1 + z K_{i}^{\beta} + O((z K_{i}^{\beta})^{2})$
for $z K_{i}^{\beta} \ll 1$. 

From Eq. (\ref{eq_mk_calc}) and 
$\partial \ln F_{i}(z) / \partial z$ 
for the above two cases of $F_{i}(z)$, 
the mean occupation number is given by 
\[
  m_{K_{i}} \approx \left\{ 
	     \begin{array}{lll}
	      z K_{i}^{\beta} & for &  z K_{i}^{\beta} \ll 1 \\
	      (z K_{i}^{\beta})^{1/\delta} & for & z K_{i}^{\beta} \geq 1.
	     \end{array} \right.
\]
Thus, 
the mean occupation number of a node increases monotonically with its
degree. 
Depending on a constant magnitude of $z$, 
we consider the following two cases (i) and (ii).

(i) We assume that the fugacity $z$ is in such a range that 
$z K_{i}^{\beta} \geq 1$ for all nodes. 
From $m_{K_{i}} = ( z K_{i}^{\beta} )^{1/\delta}$ 
for all nodes and 
the self-consistent equation 
$\rho = \frac{1}{N} \sum_{i} m_{K_{i}}
= \frac{1}{N} 
z^{1/\delta} \sum_{i} ( K_{i}^{\beta} )^{1/\delta}$, 
the solution 
$z = \rho^{\delta} / ( \bar{K}^{\beta/\delta}  )^{\delta}$ 
is valid only when 
$\bar{K}^{\beta/\delta}  \stackrel{\rm def}{=} 
\int_{K_{min}}^{K_{max}}  K^{\beta/\delta} P(K) d K$
remains finite in the limit of large $N$. 
By using $P(K) \sim K^{-\gamma}$, at $N \rightarrow \infty$, 
we have 
\[
  \int_{K_{min}}^{K_{max}}  K^{\beta/\delta} P(K) d K
  \sim \int K^{\beta/\delta - \gamma} d K
  \rightarrow K_{max}^{(\beta/\delta - \gamma + 1)}.
\]
From the negative exponent, 
this finite condition imposes that $\delta > \delta_{c}$ with 
\[
  \delta_{c} \stackrel{\rm def}{=} \frac{\beta}{\gamma -1}. 
\]
In this regime, we find 
$m_{K_{i}} \sim K_{i}^{\beta/\delta}$ 
for all nodes. 
The occupation number at the hub with the maximum degree 
$K_{max} \sim N^{1/(\gamma -1)}$ scales 
sublinearly as 
$m_{hub} \sim N^{\beta/\delta (\gamma -1)} = N^{\delta_{c}/\delta}$ 
with the exponent 
$\beta/\delta (\gamma -1) = \delta_{c} / \delta < 1$. 
Therefore, the extreme condensation of almost all packets 
at the hub with $K_{max}$ is avoided 
when $\delta > \delta_{c}$: case (A).

(ii) We assume that the fugacity $z$ is defined as 
$K_{c}^{\beta} = 1/z$ 
in the interval $K_{min} < K_{c} < K_{max}$. 
Then, the self-consistent equation becomes 
$\rho = \rho_{n} + \rho_{s} = \sum_{i} m_{K_{i}}$, where 
\begin{equation}
  \rho_{n} = \sum_{K_{min}}^{K_{c}} z K^{\beta} = K_{c}^{-\beta} 
  \int_{K_{min}}^{K_{c}} K^{\beta} P(K) d K, 
  \label{eq_rho_n}
\end{equation}
is the density for the nodes with $K < K_{c}$: 
$z K^{\beta} < 1$, and 
\begin{equation}
  \rho_{s} = \sum_{K_{c}}^{K_{max}} (z K^{\beta})^{1/\delta} 
  = K_{c}^{-\beta/\delta} 
  \int_{K_{c}}^{K_{max}} K^{\beta/\delta} P(K) d K, 
  \label{eq_rho_s}
\end{equation}
is the density for the nodes with $K > K_{c}$: 
$z K^{\beta} > 1$. 
Since 
the integral part in Eq. (\ref{eq_rho_n}) is smaller than the 
average $\beta$-power of the degree, 
$\rho_{n}$ vanishes as $K_{c}^{-\beta}$ 
in the limit of large $N$. 
In order to have a finite value of $\rho_{s} = \rho$, 
the integral part in Eq. (\ref{eq_rho_s}) 
should be divergent, which yields that 
$\delta \leq \delta_{c} = \beta/(\gamma -1)$. 
This integral should be of the same order as $K_{c}^{-\beta/\delta}$, 
which yields
\[
  K_{c} \sim \left\{ 
  \begin{array}{lll}
   (\ln K_{max})^{\delta_{c}/\beta} & for & \delta = \delta_{c} \\
   (K_{max})^{1 - \delta/\delta_{c}} & for & \delta < \delta_{c}.
  \end{array} \right.
\]
in cases (B) and (C), respectively.

\end{document}